\documentclass[aps,prl,twocolumn,preprintnumbers,superscriptaddress,dblfloatfix,nofootinbib]{revtex4-1}
\usepackage[utf8]{inputenc}
\usepackage{lmodern}
\usepackage{amsmath,amssymb}
\usepackage{graphicx}
\usepackage{url}
\usepackage{color} 
\usepackage{enumitem}
\usepackage{subfigure}
\usepackage[dvipsnames]{xcolor}
\usepackage[colorlinks=true,breaklinks=true]{hyperref}
\hypersetup{allcolors=[rgb]{0.0 0.0 0.6},linkcolor=[rgb]{0.75 0.05 0.05}}
\usepackage{bm}
\usepackage{epsfig}
\usepackage{slashed}
\usepackage{color}
\usepackage{accents}
\usepackage[dvipsnames]{xcolor}
\usepackage[colorlinks=true,breaklinks=true]{hyperref}
\hypersetup{allcolors=[rgb]{0.0 0.0 0.6},linkcolor=[rgb]{0.75 0.05 0.05}}

\renewcommand\[{\left[}

\newcommand{\exclude}[1]{}

\begin{document}
\preprint{IPMU22-0041} 

\title{ Fireball  baryogenesis from early structure formation due to Yukawa forces}
	
\author{Marcos M.  Flores} 
\affiliation{Department of Physics and Astronomy, University of California, Los Angeles \\ Los Angeles, California, 90095-1547, USA} 
\author{Alexander Kusenko} 
\affiliation{Department of Physics and Astronomy, University of California, Los Angeles \\ Los Angeles, California, 90095-1547, USA}
\affiliation{Kavli Institute for the Physics and Mathematics of the Universe (WPI), UTIAS \\The University of Tokyo, Kashiwa, Chiba 277-8583, Japan} 
\affiliation{Theoretical Physics Department, CERN, 1211 Geneva 23, Switzerland}

\author{Lauren Pearce}
\affiliation{Pennsylvania State University-New Kensington, New Kensington, Pennsylvania 15068, USA}

\author{Graham White}
\affiliation{Kavli Institute for the Physics and Mathematics of the Universe (WPI), UTIAS \\The University of Tokyo, Kashiwa, Chiba 277-8583, Japan}
\date{\today}
	
\begin{abstract}
We show that viable electroweak baryogenesis can be realized without a first-order phase transition if plasma is heated inhomogeneously by nongravitational structure formation in some particle species. Yukawa interactions can mediate relatively long-range attractive forces in the early Universe.  This creates an instability and leads to growth of structure in some species even during the radiation dominated era.  At temperatures below the electroweak scale, the collapsing and annihilating halos can heat up plasma in fireballs that expand and create the out-of-equilibrium high-temperature environment suitable for generating the baryon asymmetry.  The plasma temperature at the time of baryogenesis can be as low as a few MeV, making it consistent with both standard and low-reheat cosmologies. 

\end{abstract}
\maketitle
	

Yukawa forces are generically stronger than gravity, and they can lead to the formation of structure during both matter and radiation dominated  eras~\cite{Amendola:2017xhl,Savastano:2019zpr, Casas:2016duf,Flores:2020drq, Domenech:2021uyx, Flores:2021jas, Domenech:2023afs}. The growth of these structures can lead to bound states~\cite{Wise:2014jva,Gresham:2017cvl,Gresham:2018rqo} or, in combination with radiative cooling by the same Yukawa interactions, can lead to collapse and  formation of primordial black holes~\cite{Flores:2020drq,Flores:2021jas}. Alternatively, the halos can disappear via annihilation of their constituent particles.  However, if the halo particles have some interactions with Standard Model (SM) particles, the formation, collapse, and decay of the halos can result in local heating of the ambient plasma. 

In this work we will show that this inhomogeneous heating of plasma creates suitable conditions for electroweak baryogenesis if the halos form when the plasma temperature is below the electroweak phase transition.  The requisite departure from thermal equilibrium~\cite{Sakharov:1967dj} is achieved in the expanding fireball at the site of a halo collapse.  This allows for electroweak baryogenesis, which makes use of the  baryon-number-violating sphaleron transitions~\cite{Kuzmin:1985mm}, even in the absence of a first-order phase transition.  Our scenario is consistent with both high and extremely low reheat temperatures, broadening the range of viable cosmological models. Unlike some other models of low-scale baryogenesis~\cite{Garcia-Bellido:1999xos,Cornwall:2001hq}, we do not require preheating or any particular  inflationary scenario.  The paradigm we propose is consistent with any viable  inflationary model. This is similar to the scenario of Ref.~\cite{Asaka:2003vt}, where inhomogeneous heating of plasma was due to hadronic jets of decaying heavy particles.  

As in Ref.~\cite{Flores:2020drq}, we will consider a dark sector heavy fermion $\psi$ interacting with a scalar $\chi$ via a Yukawa interaction,
\begin{equation}
\begin{split}
\mathcal{L} &\supset
\dfrac{1}{2} \partial^\mu \chi \partial_\mu \chi
+ 
\frac{1}{2}m_\chi^2\chi^2
+
\frac{\zeta}{4!}\chi^4
\\[0.25cm]
&+ \bar{\psi} i \gamma^\mu \partial_\mu \psi - m_\psi \bar{\psi} \psi 
- y\chi\bar{\psi}\psi + \cdots
\end{split}
\end{equation}
We have included a quartic coupling, as is required for renormalizability.  We choose $\zeta \ll 1$ and neglect its effects on structure formation. In general, as shown in Ref.~\cite{Domenech:2023afs},  a quartic does not preclude the formation of structure during a radiation dominated era.

The fermions are assumed to be either stable or have a total decay width $\Gamma \ll m_\psi^2/M_{\rm Pl}$, where $M_{\rm Pl}\approx 2.4\times 10^{18}$ GeV. This allows for an era during which the $\psi$ particles become nonrelativistic and decouple from the plasma, and interact via a long-range force mediated by the $\chi$ field. Once freeze-out of $\bar{\psi}\psi\leftrightarrow \chi\chi$ interactions occurs and the mean-free path of $\chi$ particles becomes sufficiently large, $\psi$ halos can start to form and collapse.

In view of the previous discussion, halo formation begins at the growth temperature $T_{g}\equiv \min\{T_{\rm FO}, m_\psi/3\}$. The freeze-out temperature, $T_{\rm FO}$, is determined by solving $\Gamma(T) = H $, where $\Gamma$ is the interaction rate for $\bar{\psi}\psi \leftrightarrow \chi\chi$ interactions given by
\begin{equation}
\label{eq:DMIntRate}
\Gamma(T)
\simeq
\frac{y^4}{4\pi}\cdot \frac{n_\psi^{\rm EQ}(T)}{(T^2 + m_\psi^2)}	
\end{equation}
and $n_\psi^{\rm EQ}(T)$ is the  equilibrium number density for fermions.
Unlike in gravitationally bound systems, the binding energy within systems bound by long-range scalar interactions can be non-negligible. This binding energy can become a nontrivial component of the energy budget for certain values of $\{y, m_\psi, m_\chi\}$. To account for this, we include an additional energy component in the Friedmann equation,
\begin{equation}
3M_{\rm Pl}^2H^2
=
\rho_{\rm rad} + \rho_y
.
\end{equation}
where $\rho_y$ accounts for the energy density of Yukawa potential energy. The limited range of the mediator $\chi$, determined by $m_\chi^{-1}$, requires that we consider two regimes: (i) the mediator acts on all particles within the horizon ($H^{-1} < m_{\chi}^{-1}$) or (ii) the mediator can only act on subhorizon patches ($H^{-1} > m_{\chi}^{-1}$). In the latter scenario, the number of regions subject to scalar interactions within the horizon is $N_h = (m_\chi/H)^3$.

The above two scenarios lead to two possible expressions for the Yukawa energy density,
\begin{equation}
\rho_y(T)
=
\frac{3y^2}{4\pi m_\psi^2 H^{-3}}
\begin{cases}
M_{\rm hor}^2/H^{-1} & H^{-1} < m_\chi^{-1}\\
N_h M_{\rm hal}^2/m_\chi^{-1} & H^{-1} > m_\chi^{-1},
\end{cases}
\end{equation}
where
\begin{equation}
\begin{pmatrix}
M_{\rm hor}\\
M_{\rm hal}
\end{pmatrix}
=
\frac{4\pi}{3}m_\psi n_{\psi}^{\rm eq}(T)
\begin{pmatrix}
H(T)^{-3}\\
m_\chi^{-3}
\end{pmatrix}.
\end{equation}

The three parameters $\{m_\psi, y, m_\chi\}$, establish three important temperatures. The Universe originally begins in a radiation dominated epoch and transitions to Yukawa domination at $T_{\rm eq}^{\rm RD\to \rm YD}$. The horizon size continues to grow until $H^{-1}$ exceeds $m_{\chi}^{-1}$ at $T_{m_\chi = H}$. Lastly, the number density of the $\psi$ fluid rapidly decreases as the temperature falls below $m_\psi$. This allows for reestablishment of radiation domination at $T_{\rm eq}^{\rm YD\to  RD}$. Taking this into account, the evolution of the Hubble parameter becomes
\begin{equation}
H(T)^2
=
\begin{cases}
	\frac{\pi^2}{90}g_*\frac{T^4}{M_{\rm Pl}^2} &  T \lesssim T_{\rm eq}^{\rm YD\to \rm RD}\ \&\ T \gtrsim T_{\rm eq}^{\rm RD\to YD}\\[0.25cm]
	\frac{2\pi^{1/2}}{3M_{\rm Pl}}yn_\psi^{\rm eq}(T) & T_{m_\chi = H} \lesssim T \lesssim T_{\rm eq}^{\rm RD\to YD}\\[0.25cm]
	\frac{4\pi}{9M_{\rm Pl}^2}\frac{y^2n_{\psi}^{\rm eq}(T)^2}{m_\chi^2} & T_{\rm eq}^{\rm YD\to  RD} \lesssim T \lesssim T_{m_\chi = H}.
	\end{cases}	
\end{equation}
Depending on the parameters, radiation domination may be reestablished before the horizon exceeds the Compton wavelength of the mediator. In this case, the evolution of the Hubble parameter is instead given by
\begin{equation}
H(T)^2
=
\begin{cases}
	\frac{\pi^2}{90}g_*\frac{T^4}{M_{\rm Pl}^2} &  T \lesssim T_{\rm eq}^{\rm YD\to RD}\ \&\ T \gtrsim T_{\rm eq}^{\rm RD\to YD}\\[0.25cm]
	\frac{2\pi^{1/2}}{3M_{\rm Pl}}yn_\psi^{\rm eq}(T) & T_{\rm eq}^{\rm YD\to RD} \lesssim T \lesssim T_{\rm eq}^{\rm RD\to \rm YD}.
\end{cases}	
\end{equation}
Given the Hubble parameter in this more general framework, determining when the $\bar{\psi}\psi\leftrightarrow \chi\chi$ freeze-out is a simple matter of computing the temperature, $T_{\rm FO}$, where $\Gamma(T_{\rm FO})/H(_{\rm FO}) = 1$, where $\Gamma(T)$ is given by Eq. \eqref{eq:DMIntRate}.

Given $T_{\rm FO}$ and therefore the growth temperature $T_g$, we can estimate the properties of the $\psi$ halos formed as a result of long-range interactions. The halo mass is  approximated by the mass enclosed within the radius $R_h = m_\chi^{-1}$ at the growth temperature,
\begin{equation}
M_h 
=
\frac{4\pi}{3}m_\psi n_\psi(T_g)R_h^3	
.
\end{equation}

In the absence of dissipation, these dark matter halos would remain as such until their constituent particles decayed. However, the same Yukawa interaction introduces a dissipation channel for energy and angular momentum through emission of scalar radiation. Scalar cooling proceeds through different emission channels depending on the size and density of the $\psi$ halo. For dilute systems, incoherent oscillatory motion of $\psi$ particles leads to emission power $P\propto y^2\omega^4R^2M_h$. For halos with higher densities, pairwise interactions lead to scalar bremsstrahlung similar to free-free emission of photons from plasma. The final stage of collapse begins with $\chi$ radiation becoming trapped, leading to surface emission from a fireball configuration of the halo. This stage sets the timescale for collapse since surface radiation is by far the least efficient at removing energy from the halo. In particular, for a halo of radius $R$, $\tau_{\rm col}\sim R \ll H^{-1}$~\cite{Flores:2020drq}. Therefore, we take the temperature of the plasma outside of the halo to be the growth temperature $T_{g}$. 

The final state of $\psi$ halos depends largely on their size and their ability to overcome Fermi pressure. If the Compton wavelength of the heavy fermions fits inside the Schwarzschild radius of their respective halos,

\begin{equation}
\frac{1}{m_\psi} < \frac{M_h}{4\pi M_{\rm Pl}^2}
,
\end{equation}
then the halo will collapse into a black hole.  However, this condition is not satisfied for the parameter space we consider in this study.  Instead, the halos collapse until the average particle spacing within the halo is comparable to $1/m_\psi$. At this point, the halos annihilate as there is no asymmetry in the $\psi$ population.  

We note that thermal corrections to the $\chi$ mass might impede the formation or collapse of $\psi$ halos. First, coupling the scalar $\chi$ to the SM might induce thermal corrections due to the ambient SM plasma. In what follows, we will directly couple $\chi$ to the Higgs sector.  During the formation and collapse of the dark sector halos, the SM background temperature always remains below $T\sim 100$ GeV. Therefore, temperature corrections arising from the direct coupling of the $\chi$ particle to the Higgs sector will be Boltzmann suppressed, i.e., $\delta m_\chi \propto \exp(-m_H/T)$. Thus, we do not expect the SM plasma to offer any major corrections to the scalar mass $m_\chi$.  Inside the collapsed halo, the temperature increases beyond 100~GeV, and the  temperature corrections to $m_\chi$ may be significant, but as the halo has already collapsed and transferred energy to the SM sector, these corrections are unimportant.

Second, temperature corrections from the dark sector may come into play in the later stages of halo collapse. Eventually, the scalar radiation becomes trapped in a collapsing halo, resulting in a correction to $m_\chi$ due to the local $\psi$ density. A decrease in the range of the scalar interaction does not necessarily affect the halo stability, in analogy with large atomic nuclei which are maintained by attractive interactions whose range is much smaller than their physical size.

The final stages of halo collapse are complicated by bound state formation, convection, etc. We  leave these considerations for future work.

Regardless of the final state of dark matter halo, the energy released from collapse is~\cite{Flores:2020drq}
\begin{equation}
\Delta E_{\rm col} = \frac{y^2M_h^2}{m_\psi^2 R_c}
\left(1- 
\frac{R_c}{R_h} 
\right),
\end{equation}
where $R_c$ is the critical radius which is either the radius where annihilations become significant or the Schwarzschild radius.
The annihilating halos also release energy into the ambient plasma, $\Delta E_{\rm ann} = \epsilon_{\rm ann} M_h$, where $\epsilon_{\rm ann}$ parametrizes the efficiency of annihilation. The total energy release is the sum of these two contributions, $\Delta E \equiv \Delta E_{\rm col} + \Delta E_{\rm ann}$. 
We assume that the dark sector consisting of $\psi$ and $\chi$ is weakly coupled to the SM, and so the sudden release of this large quantity of energy locally heats the plasma. In particular, immediately after collapse and annihilation of the halo the temperature of the heated region is 
\begin{equation}
T_i^4
=
\frac{90\xi_s\Delta E}{4\pi^3 g_*(T_i)R_i^3},
\label{eq:Ti}
\end{equation}
where $R_i$ is the initial radius of the heated region, $g_*(T_i)$ are the relativistic degrees of freedom at $T_i$ and, most importantly, $\xi_s$ is the efficiency of energy transfer of energy from the dark sector $\chi$ particles to the SM plasma.

The specifics of the efficiency depend on the specific coupling of the dark sector to the SM. One may consider, for example, the quartic coupling $\mathcal{L}\supset \lambda
\ \chi\chi H H$. In this case, the coupling $\lambda$ and $\xi_s$ are related through
\begin{align}
&\lambda^2
=
\frac{4\pi^3}{\zeta(3)}
\left(
\frac{T_{\rm eff} m_\psi}{g_*(T_i)T_{\rm bg}^2 }
\right)\\[0.25cm]
&\simeq
\frac{\xi_s}{100}
\left(
\frac{106.75}{g_*(T_i)}
\right)
\Bigg(
\frac{T_{\rm eff}}{100\ {\rm GeV}}
\Bigg)
\Bigg(
\frac{m_\psi}{100\ {\rm GeV}}
\Bigg)
\Bigg(
\frac{100\ {\rm GeV}}{T_{\rm bg}}
\Bigg)^2\nonumber,
\end{align}
where $T_{\rm eff}$ is the effective temperature of the halo given by
\begin{equation}
T_{\rm eff} = \left(\frac{y^2 M_h}{m_\psi R^4}\right)^{1/4}
.
\end{equation}
For all of the parameter space we consider, $\lambda \ll 0.1$ and therefore it is unimportant in our later discussion of \textit{CP} violation.

Once a region is heated above the background temperature, dissipation of the fireball can occur via two main channels. First, the energy release may be very rapid, via an expanding shock wave which travels through the plasma at the speed of sound. Utilizing conservation of energy, we define the characteristic timescale associated with this expansion as
\begin{equation}
\tau_{\rm exp}
\equiv \frac{T}{|dT/dt|}
=
\frac{4R_i}{\sqrt{3}}
\left(
1 + \frac{t - t_i}{\sqrt{3}R_i}
\right),
\end{equation}
where $t_i$ is the time when the heated region formed. Since most of the energy is released at the end of collapse, we approximate $R_{i} = \max\{2GM_h, R_{\rm ann}\}$, where $R_{\rm ann}$ is the halo radius in which the average distance between $\psi$ particles is $m_\chi^{-1}$ and is explicitly given by
\begin{equation}
R_{\rm }
=
m_\psi^{-1}
\left(
\frac{3}{4\pi}
\frac{M_h}{m_\psi}
\right)
.
\end{equation}

Alternatively, the fireball might cool through diffusion on the timescale~\cite{Asaka:2003vt}
\begin{equation}
\tau_{\rm diff}
\sim
\frac{R_i^2}{4D}
\left(
\frac{T_i}{T}
\right)^{8/3},
\end{equation}
where $D$ is a diffusion constant. As in Ref.~\cite{Asaka:2003vt}, we take $D \sim 1/\gamma_g$ where $\gamma_g\sim 0.3 g_s^2 T$ and $g_s$ is the strong coupling~\cite{Braaten:1990it}. We may then express the diffusion timescale as
\begin{equation}
\tau_{\rm diff}
\sim
\frac{3}{40}g_s^2(T)T
\left(
\frac{90}{4\pi^3}
\frac{\xi_s|\Delta E|}{g_*(T_i)T^4}
\right)^{2/3}
.
\end{equation}
The dissipation timescale is defined as $\tau_{\rm diss} \equiv \min\{\tau_{\rm exp}, \tau_{\rm diff}\}$. Generally, $\tau_{\rm exp} \ll \tau_{\rm diff}\propto T_i^{8/3}$ which can be significantly larger than the temperature of the heated region soon after the beginning of its evolution. 

The temperature in Eq.~\eqref{eq:Ti} is generally above the electroweak scale.  Thus, the SM plasma undergoes two phase transitions: a rapid one as it initially  heats, and another one as it cools.  This provides a natural environment for electroweak baryogenesis, even if the postinflationary reheating temperature is below the electroweak scale.  In this scenario, only a fraction of the SM plasma is heated, naturally leading to a small baryon asymmetry.

{\it Example of a specific model that illustrates the new paradigm.}   Although we do not need a first order phase transition, we do need additional \textit{CP} violation beyond that present in the SM. In principle, any electroweak baryogenesis scenario is applicable here. As an example, we consider a well-motivated scenario, a two Higgs doublet model, with the tree-level potential 
\begin{align}
V_{\mathrm{tree}} &=  
\lambda_1 
\left( H_1^\dagger H_1 -\frac{v_1^2}{2}\right)^2
+\lambda_2 \left(H_2^\dagger H_2-\frac{v_2^2}{2}\right)^2 \nonumber \\
& \qquad 
+\lambda_3 \left(\left( H_1^\dagger H_1 -\frac{v_1^2}{2}\right)+\left(H_2^\dagger H_2-\frac{v_2^2}{2}\right)\right)^2 \nonumber \\
& \qquad 
+ \lambda_5 \left( {\rm Re}(H_1^\dagger H_2) -\frac{1}{2} \cos (\xi ) (v_1 v_2 )\right)^2 \nonumber \\
& \qquad + \lambda_6 \left({\rm Im}(H_1^\dagger H_2)-\frac{1}{2} \sin (\xi ) (v_1 v_2)\right)^2.
\label{eq:tree_potential}
\end{align}
The fields acquire vacuum expectation values $\phi_1 e^{i \theta_1} \slash  \sqrt{2}$ and $\phi_2 e^{i \theta_2} \slash \sqrt{2}$.  We note that the potential (although not the Lagrangian) is invariant under the transformation $H_1 \rightarrow H_1 e^{i \zeta}$, $H_2 \rightarrow H_2 e^{i \zeta}$.  Therefore, it depends only on the difference of the phases $\theta = \theta_1 - \theta_2$.  Specific values of the coupling constants are given in the Appendix; they were chosen so that when the potential $V_{\rm tree} + V_{\rm 1-loop}$ is minimized, the lowest mass eigenstate and the vacuum expectation value (VEV) $\sqrt{ \phi_1^2 + \phi_2^2}$ match observed values.  The chosen parameters also have $\tan\beta = 2.68$ and $\beta - \alpha = 1.63$, consistent with observational constraints~\cite{Chowdhury:2017aav}.  Since $\xi = 0$, there is no additional \textit{CP} violation at low temperature, allowing us to bypass constraints from negative searches for electric dipole moments. We also can avoid flavor changing neutral currents by coupling only one doublet to fermions. In the Appendix, we give expressions for $V_{\rm 1-loop}$ as well as the finite temperature corrections $V_T$.  

The above example is by no means unique, and there are many other possibilities for coupling $\chi$ to the SM and for the sources of \textit{CP} violation.

\begin{figure}
\centering
\includegraphics[width=0.48\textwidth]{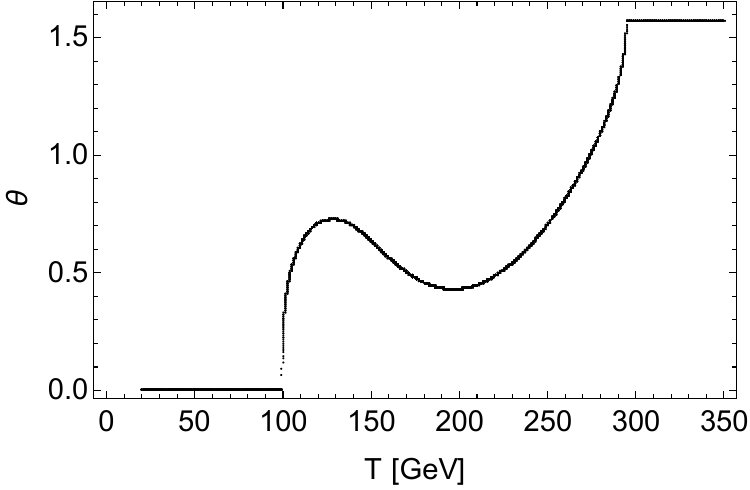}
\caption{The evolution of the value of $\theta$ in the minimum of the potential, as a function of temperature.  Although $\theta$ evolves rapidly at times, the smooth evolution indicates a second order phase transition.  
}
\label{fig:theta}
\end{figure}

At sufficiently high temperatures, the minimum of the potential is at $\theta = \pi \slash 2$.  As the plasma cools, it undergoes a smooth transition to a minimum with $\theta = 0$, as shown in Fig.~\ref{fig:theta}.  Because of the symmetry mentioned above, we expect $\theta_1 + \theta_2$ to remain constant as long as the phase transition is slow enough that $\theta_-$ remains in the value that minimizes the potential.  Therefore, $\dot{\theta}_1 = - \dot{\theta}_2 = \dot{\theta} \slash 2$.

To avoid flavor changing neutral currents, we consider here a ``type I" two Higgs doublet model~\cite{Bhattacharyya:2015nca}, in which all fermions are coupled to $H_2$.  Consequently, after spontaneous symmetry breaking the Yukawa terms acquire the phase $e^{i \theta_2}$.  This can be removed via a rotation of either the left- or right-handed fermion, but because $\theta_2$ is time dependent, the Lagrangian acquires a $\dot{\theta}_i J_B^0$ term (where $J_{B+L}^0$ is the baryon number density), from the fermion kinetic energy terms.  This can be generalized to types II, X, and Y two Higgs doublet models.

This term is analogous to ones that appear in spontaneous baryogenesis models and, in the presence of a baryon-number-violating process, an asymmetry is produced~\cite{Dine:1990fj,Cohen:1990it,Dolgov:1996qq,Dolgov:1997qr,Cornwall:2001hq,Kusenko:2014lra,Kusenko:2014uta,Yang:2015ida}.  During the second order phase transition when $\theta_2$ evolves slowly, it may be possible to treat it as a background field, while the plasma remains in approximate thermal equilibrium biased by an effective chemical potential~\cite{Dolgov:1997qr,Yang:2015ida}:
\begin{equation}
\mu_{\rm eff} =  \dfrac{\dot{\theta} }{2}
= - \dfrac{1}{2} \dfrac{T}{\tau_{\rm diss}} \dfrac{d\theta}{dT},
\end{equation}
where $dT \slash dt \approx -T \slash \tau_{\rm diss}$ measures the rate of change of temperature. 
In our case, the second-order phase transition takes place in an expanding fireball, in which the departure from thermal equilibrium~\cite{Sakharov:1967dj} is much more dramatic than in the case of a cosmological phase transition ($\tau_{\rm diss}\ll H^{-1}$). 
We note that as shown in Fig.~\ref{fig:theta}, $d\theta \slash dT$ and thus the effective potential changes sign at temperatures around $125$ and $200 \, \mathrm{GeV}$. 
The effective chemical potential leads to a baryon asymmetry only in the presence of baryon-number-violating processes; during the electroweak phase transition, sphalerons are such a process.  The evolution of baryonic number density in these heated regions is described by 
\begin{equation}
\frac{d n_B(T)}{d \ln T}
=
\frac{\Gamma_{\rm sph}(T)}{\tau_{\rm diss}^{-1}(T)}
\left(
n_B(T) - \mu_{\rm eff} T^2
\right)
\end{equation}
while the fireball cools.  This equation, which is similar to  the Boltzmann equation in traditional freeze-out calculations, demonstrates that at some temperature $T_{f}$, the baryonic number density was frozen into
\begin{equation}
n_B
\simeq
\left.
\mu_{\rm eff} T^2
\right|_{T = T_f},
\end{equation}
where the temperature $T_f$ is approximated by solving $\Gamma_{\rm sph}(T) = \tau_{\rm diss}^{-1}(T)$. For the sphaleron rate we use an expression that reduces to the lattice result at small $v/T$,
\begin{equation}
    \Gamma _{\rm sph}(T) = 6\kappa \alpha ^5 T e^{-\Delta E_{\rm sph}/T}
\end{equation}
with $\kappa =20$ \cite{Bodeker:1999gx,Moore:1999fs,Moore:2000mx}, as calculations of the fluctuation determinant have only been performed in the limit $v/T \gg 0.1$ \cite{Carson:1990jm, Baacke:1993aj}. We note that more sophisticated estimations of the sphaleron rate prefactor exist, i.e., Ref.~\cite{DOnofrio:2014rug}. These modifications will at best offer logarithmic corrections to the exponential above, but should be included in a more complete analysis.

We approximate the resultant baryon asymmetry as
\begin{equation}
\eta_B
\sim
f\frac{n_B}{T_{\rm bg}^3}, 
\end{equation}
where $f$ is the volume filling factor defined as
\begin{equation}
f\equiv N_h H^3(T_{\rm bg})R_i^3
\left(
\frac{T_i}{T_f}
\right)^{4},
\end{equation}
and $T_{\rm bg}$ is the background temperature, approximated as $T_{g}$. $f$ measures the fraction of the Hubble volume in the superheated bubbles when the sphalerons freeze out.
The baryon asymmetry observed today is related to baryon asymmetry produced  through
\begin{equation}
\frac{\eta_{B}(T_0)}{\eta_B(T_{\rm bg})}
=
\frac{g_{*, S}(T_{\rm 0})}{g_{*, S}(T_{\rm bg})}
\sim
0.04
\end{equation}
where $T_0$ is the present day temperature of the Universe and $g_{*, S}$ is the conventional relativistic degrees of freedom relevant to entropy.

The baryon asymmetry is determined by four free parameters $\{m_\psi, y, m_\chi, \xi_s\}$. There are numerous restrictions on these which we have enforced to provide a conservative exploration of the parameter space: {\it (i)} We require that $0.01\ {\rm GeV} \lesssim T_{\rm FO} \lesssim 100$ GeV, as is consistent with our initial assumption of cold EW baryogenesis, {\it (ii)} $f < 1$ as is necessary, {\it (iii)} $T_i < M_{\rm Pl}$, and {\it (iv)} $T_i > 350$ GeV. This final condition ensures that $\theta(t)$ is evolving as inhomogenously heated regions begin to cool.

\begin{figure}
\centering
\includegraphics[width=0.48\textwidth]{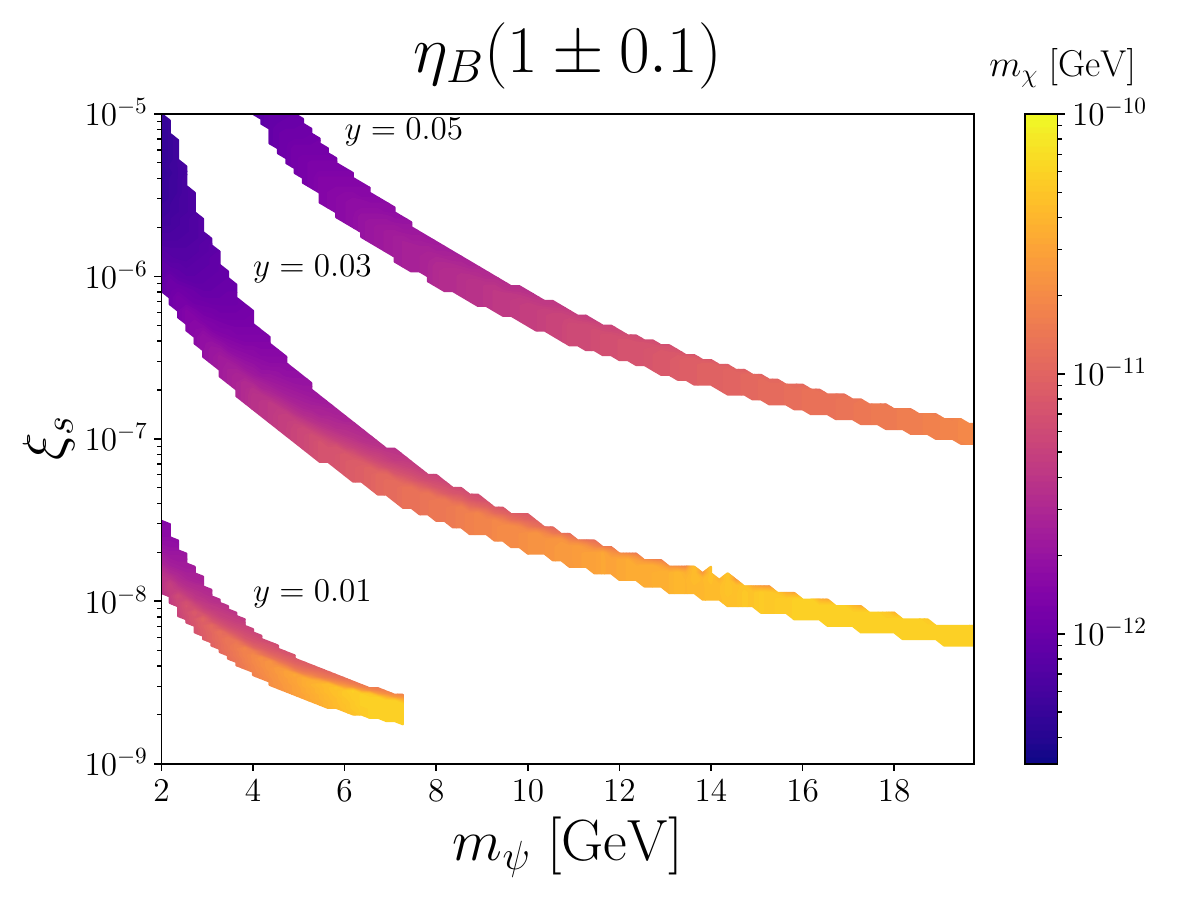}
\caption{Regions of parameter space which produce a baryon asymmetry consistent within 10\% of the observed value $\eta_B =  6.129\times 10^{-10}$~\cite{Planck:2018vyg}. Each band represents fixed values of the coupling $y$, while the colors indicate varying values of the mediator mass $m_\chi$.}
\label{fig:etaB}
\end{figure}

Figure~\ref{fig:etaB} illustrates the regions of parameter space which produce baryon asymmetries within 10\% of the observed value, $\eta_B = 6.129\times 10^{-10}$~\cite{Planck:2018vyg}. Each band represents a fixed value of the coupling, while the contours correspond to different values of the mediator mass $m_\chi$.

Depending on the temperature at which the sphaleron transition decouples, the time derivative of the phase can be positive or negative (see Fig.~\ref{fig:theta}).  The temperature is determined by the  fireball expansion rate, which  depends on the size of the dark halo.   The distribution of halos is model dependent.  While it is possible that some fireballs have positive and some have negative signs of $\dot \theta$ at the time of the sphaleron decoupling, no significant cancellation is expected.  The overall asymmetry per Hubble volume is a convolution of the $\dot \theta (T)$ with a function that depends on the  distribution of the halo sizes.  Since these two functions are independent, it is implausible that the convolution integral might vanish.  The overall sign of the baryon asymmetry is subject to convention and redefinition through the relative phases of the Higgs doublets. 

If domains with different values and signs of the asymmetry appear from different size halos, the averaging takes place on the timescales associated with diffusion. If one assumes the halo separation distance of the order of $L \sim f^{1/3} H_{\rm bg}^{-1} \sim f^{1/3} M_{\rm Pl}/T_{\rm bg}^2$, any inhomogeneities in the  distribution of the baryon asymmetry average out on the timescale $ L^2/4D_{\rm B}\sim  (L^2 g^2 T/40)$~\cite{Arnold:2000dr} at temperature $T$.  This timescale equals the horizon time $H^{-1}=M_{\rm Pl}/T^2$ for   $T \sim (40\, T_{\rm bg}^4/f^{2/3} g^2 M_{\rm Pl})^{1/3} > {\rm MeV}$ for $T_{\rm bg}\sim 70$~GeV and $f <0.1$.  For lower temperatures, big-bang nucleosynthesis may be inhomogeneous~\cite{Jedamzik:2001qc,Matsuura:2004ss,Dolgov:2008wu,Nakamura:2017qtu}. 

The presence of an extra relativistic degree of freedom, in the form of $\chi$, also suggests a slightly different value for the effective number of degrees of freedom, $N_{\rm eff}$. In the dark sector the effective degrees of freedom is reduced by a factor of $2\times(7/8)$ as the $\psi$ population becomes nonrelativistic. Assuming the SM and dark sector have the same temperature around $T\sim m_\psi$ then the contribution to the measured value of $N_{\rm eff}$ is~\cite{Blennow:2012de,Fuller:2011qy,Patwardhan:2015kga}
\begin{equation}
\Delta N_{\rm eff}
\approx
14[g_1/g_*(T_d)]^{4/3}
\approx 0.1 - 0.2,
\end{equation}
where $T_d\sim 1 - 100$ GeV is the temperature where the dark sector and SM decoupled and $g_1$ are the dark sector degrees of freedom before decoupling. Not only is  $\Delta N_{\rm eff}\sim 0.2$ allowed, but it may help alleviate the tension between different measurements of the Hubble constant~\cite{Bernal:2016gxb,Gelmini:2019deq,Anchordoqui:2019yzc,Vattis:2019efj,Escudero:2019gvw,Gelmini:2020ekg,Vagnozzi:2019ezj,Wong:2019kwg, Perivolaropoulos:2021jda}.

Fireball baryogenesis is consistent with both high and low reheat temperatures, limited only by the viability of nucleosynthesis.  Observational data allow the reheat temperature after inflation to  be as low as a few MeV~\cite{Kawasaki:2000en,Hannestad:2004px,Gelmini:2004ah,Gelmini:2006mr,Grin:2007yg,Gelmini:2008fq,Gelmini:2008ti,Rehagen:2015zma}.  Such nonstandard cosmology is only viable if the baryon asymmetry of the universe can be generated at a very low temperature.  While some models can generate the baryon asymmetry below the electroweak scale~\cite{Affleck:1984fy,Dine:2003ax,Garcia-Bellido:1999xos,Cornwall:2001hq,Asaka:2003vt, Elor:2020tkc, Elahi:2021jia, Jaeckel:2022osh}, not all of them can work below 10~GeV.  Our model is viable even at $T_{\rm bg}\sim 10$~MeV, broadening the range of nonstandard cosmologies.

The model has interesting phenomenological implications.   Figure~\ref{fig:etaB} shows the baryon asymmetry can be generated with sizable energy exchange between the visible and dark sectors, parametrized by $\xi_s$.  While this exchange can be mediated at any scale up to $T_i \sim 10^{11} \, \mathrm{GeV}$, lower scales may be probed experimentally~\cite{Batell:2009di,Kanemura:2010sh,Fichet:2017bng,Contino:2020tix}.

In summary, we have shown that primordial structure formation due to the Yukawa forces can cause  inhomogeneous heating of plasma leading to baryogenesis in the presence of some sources of \textit{CP} violation.  The departure from thermal equilibrium in expanding fireballs at the sites of dark halos is sufficient for generating the baryon asymmetry of the Universe.  We illustrated the paradigm on a model with two Higgs doublets, which were used for a source of \textit{CP} violation.  One can imagine other  models that utilize the same out-of-equilibrium scenario but use different sources of \textit{CP} violation.  This opens multiple new avenues for electroweak baryogenesis without a first-order phase transition.

\begin{acknowledgments}
We thank D.~Inman, G.~Dom\`enech, and M.~Sasaki for discussions. 
A.K. was supported by the U.S. Department of Energy (DOE) Grant No. DE-SC0009937, by the Simons Foundation, by the World Premier International Research Center Initiative (WPI), MEXT, Japan, by Japan Society for the Promotion of Science (JSPS) KAKENHI Grant No. JP20H05853, and by the UC Southern California Hub with funding from the UC National Laboratories division of the University of California Office of the President. The work of G.W. is supported by World Premier International Research
Center Initiative (WPI), MEXT, Japan. G.W. was supported by JSPS KAKENHI Grant No. JP22K14033. This work used computational and storage services associated with the Hoffman2 Shared Cluster provided by UCLA Institute for Digital Research and Education’s Research Technology Group.
\end{acknowledgments}


\bibliography{biblio}

 \newcommand{\noop}[1]{}
\begin{thebibliography}{69}%
\makeatletter
\providecommand \@ifxundefined [1]{%
 \@ifx{#1\undefined}
}%
\providecommand \@ifnum [1]{%
 \ifnum #1\expandafter \@firstoftwo
 \else \expandafter \@secondoftwo
 \fi
}%
\providecommand \@ifx [1]{%
 \ifx #1\expandafter \@firstoftwo
 \else \expandafter \@secondoftwo
 \fi
}%
\providecommand \natexlab [1]{#1}%
\providecommand \enquote  [1]{``#1''}%
\providecommand \bibnamefont  [1]{#1}%
\providecommand \bibfnamefont [1]{#1}%
\providecommand \citenamefont [1]{#1}%
\providecommand \href@noop [0]{\@secondoftwo}%
\providecommand \href [0]{\begingroup \@sanitize@url \@href}%
\providecommand \@href[1]{\@@startlink{#1}\@@href}%
\providecommand \@@href[1]{\endgroup#1\@@endlink}%
\providecommand \@sanitize@url [0]{\catcode `\\12\catcode `\$12\catcode
  `\&12\catcode `\#12\catcode `\^12\catcode `\_12\catcode `\%12\relax}%
\providecommand \@@startlink[1]{}%
\providecommand \@@endlink[0]{}%
\providecommand \url  [0]{\begingroup\@sanitize@url \@url }%
\providecommand \@url [1]{\endgroup\@href {#1}{\urlprefix }}%
\providecommand \urlprefix  [0]{URL }%
\providecommand \Eprint [0]{\href }%
\providecommand \doibase [0]{http://dx.doi.org/}%
\providecommand \selectlanguage [0]{\@gobble}%
\providecommand \bibinfo  [0]{\@secondoftwo}%
\providecommand \bibfield  [0]{\@secondoftwo}%
\providecommand \translation [1]{[#1]}%
\providecommand \BibitemOpen [0]{}%
\providecommand \bibitemStop [0]{}%
\providecommand \bibitemNoStop [0]{.\EOS\space}%
\providecommand \EOS [0]{\spacefactor3000\relax}%
\providecommand \BibitemShut  [1]{\csname bibitem#1\endcsname}%
\let\auto@bib@innerbib\@empty
\bibitem [{\citenamefont {Amendola}\ \emph {et~al.}(2018)\citenamefont
  {Amendola}, \citenamefont {Rubio},\ and\ \citenamefont
  {Wetterich}}]{Amendola:2017xhl}%
  \BibitemOpen
  \bibfield  {author} {\bibinfo {author} {\bibfnamefont {L.}~\bibnamefont
  {Amendola}}, \bibinfo {author} {\bibfnamefont {J.}~\bibnamefont {Rubio}}, \
  and\ \bibinfo {author} {\bibfnamefont {C.}~\bibnamefont {Wetterich}},\ }\href
  {\doibase 10.1103/PhysRevD.97.081302} {\bibfield  {journal} {\bibinfo
  {journal} {Phys. Rev. D}\ }\textbf {\bibinfo {volume} {97}},\ \bibinfo
  {pages} {081302} (\bibinfo {year} {2018})},\ \Eprint
  {http://arxiv.org/abs/1711.09915} {arXiv:1711.09915 [astro-ph.CO]}
  \BibitemShut {NoStop}%
\bibitem [{\citenamefont {Savastano}\ \emph {et~al.}(2019)\citenamefont
  {Savastano}, \citenamefont {Amendola}, \citenamefont {Rubio},\ and\
  \citenamefont {Wetterich}}]{Savastano:2019zpr}%
  \BibitemOpen
  \bibfield  {author} {\bibinfo {author} {\bibfnamefont {S.}~\bibnamefont
  {Savastano}}, \bibinfo {author} {\bibfnamefont {L.}~\bibnamefont {Amendola}},
  \bibinfo {author} {\bibfnamefont {J.}~\bibnamefont {Rubio}}, \ and\ \bibinfo
  {author} {\bibfnamefont {C.}~\bibnamefont {Wetterich}},\ }\href {\doibase
  10.1103/PhysRevD.100.083518} {\bibfield  {journal} {\bibinfo  {journal}
  {Phys. Rev. D}\ }\textbf {\bibinfo {volume} {100}},\ \bibinfo {pages}
  {083518} (\bibinfo {year} {2019})},\ \Eprint
  {http://arxiv.org/abs/1906.05300} {arXiv:1906.05300 [astro-ph.CO]}
  \BibitemShut {NoStop}%
\bibitem [{\citenamefont {Casas}\ \emph {et~al.}(2016)\citenamefont {Casas},
  \citenamefont {Pettorino},\ and\ \citenamefont {Wetterich}}]{Casas:2016duf}%
  \BibitemOpen
  \bibfield  {author} {\bibinfo {author} {\bibfnamefont {S.}~\bibnamefont
  {Casas}}, \bibinfo {author} {\bibfnamefont {V.}~\bibnamefont {Pettorino}}, \
  and\ \bibinfo {author} {\bibfnamefont {C.}~\bibnamefont {Wetterich}},\ }\href
  {\doibase 10.1103/PhysRevD.94.103518} {\bibfield  {journal} {\bibinfo
  {journal} {Phys. Rev. D}\ }\textbf {\bibinfo {volume} {94}},\ \bibinfo
  {pages} {103518} (\bibinfo {year} {2016})},\ \Eprint
  {http://arxiv.org/abs/1608.02358} {arXiv:1608.02358 [astro-ph.CO]}
  \BibitemShut {NoStop}%
\bibitem [{\citenamefont {Flores}\ and\ \citenamefont
  {Kusenko}(2021)}]{Flores:2020drq}%
  \BibitemOpen
  \bibfield  {author} {\bibinfo {author} {\bibfnamefont {M.~M.}\ \bibnamefont
  {Flores}}\ and\ \bibinfo {author} {\bibfnamefont {A.}~\bibnamefont
  {Kusenko}},\ }\href {\doibase 10.1103/PhysRevLett.126.041101} {\bibfield
  {journal} {\bibinfo  {journal} {Phys. Rev. Lett.}\ }\textbf {\bibinfo
  {volume} {126}},\ \bibinfo {pages} {041101} (\bibinfo {year} {2021})},\
  \Eprint {http://arxiv.org/abs/2008.12456} {arXiv:2008.12456 [astro-ph.CO]}
  \BibitemShut {NoStop}%
\bibitem [{\citenamefont {Dom\`enech}\ and\ \citenamefont
  {Sasaki}(2021)}]{Domenech:2021uyx}%
  \BibitemOpen
  \bibfield  {author} {\bibinfo {author} {\bibfnamefont {G.}~\bibnamefont
  {Dom\`enech}}\ and\ \bibinfo {author} {\bibfnamefont {M.}~\bibnamefont
  {Sasaki}},\ }\href {\doibase 10.1088/1475-7516/2021/06/030} {\bibfield
  {journal} {\bibinfo  {journal} {JCAP}\ }\textbf {\bibinfo {volume} {06}},\
  \bibinfo {pages} {030} (\bibinfo {year} {2021})},\ \Eprint
  {http://arxiv.org/abs/2104.05271} {arXiv:2104.05271 [hep-th]} \BibitemShut
  {NoStop}%
\bibitem [{\citenamefont {Flores}\ and\ \citenamefont
  {Kusenko}(2023)}]{Flores:2021jas}%
  \BibitemOpen
  \bibfield  {author} {\bibinfo {author} {\bibfnamefont {M.~M.}\ \bibnamefont
  {Flores}}\ and\ \bibinfo {author} {\bibfnamefont {A.}~\bibnamefont
  {Kusenko}},\ }\href {\doibase 10.1088/1475-7516/2023/05/013} {\bibfield
  {journal} {\bibinfo  {journal} {JCAP}\ }\textbf {\bibinfo {volume} {05}},\
  \bibinfo {pages} {013} (\bibinfo {year} {2023})},\ \Eprint
  {http://arxiv.org/abs/2108.08416} {arXiv:2108.08416 [hep-ph]} \BibitemShut
  {NoStop}%
\bibitem [{\citenamefont {Dom\`enech}\ \emph {et~al.}(2023)\citenamefont
  {Dom\`enech}, \citenamefont {Inman}, \citenamefont {Kusenko},\ and\
  \citenamefont {Sasaki}}]{Domenech:2023afs}%
  \BibitemOpen
  \bibfield  {author} {\bibinfo {author} {\bibfnamefont {G.}~\bibnamefont
  {Dom\`enech}}, \bibinfo {author} {\bibfnamefont {D.}~\bibnamefont {Inman}},
  \bibinfo {author} {\bibfnamefont {A.}~\bibnamefont {Kusenko}}, \ and\
  \bibinfo {author} {\bibfnamefont {M.}~\bibnamefont {Sasaki}},\ }\href@noop {}
  {\  (\bibinfo {year} {2023})},\ \Eprint {http://arxiv.org/abs/2304.13053}
  {arXiv:2304.13053 [astro-ph.CO]} \BibitemShut {NoStop}%
\bibitem [{\citenamefont {Wise}\ and\ \citenamefont
  {Zhang}(2014)}]{Wise:2014jva}%
  \BibitemOpen
  \bibfield  {author} {\bibinfo {author} {\bibfnamefont {M.~B.}\ \bibnamefont
  {Wise}}\ and\ \bibinfo {author} {\bibfnamefont {Y.}~\bibnamefont {Zhang}},\
  }\href {\doibase 10.1103/PhysRevD.90.055030} {\bibfield  {journal} {\bibinfo
  {journal} {Phys. Rev. D}\ }\textbf {\bibinfo {volume} {90}},\ \bibinfo
  {pages} {055030} (\bibinfo {year} {2014})},\ \bibinfo {note} {[Erratum:
  Phys.Rev.D 91, 039907 (2015)]},\ \Eprint {http://arxiv.org/abs/1407.4121}
  {arXiv:1407.4121 [hep-ph]} \BibitemShut {NoStop}%
\bibitem [{\citenamefont {Gresham}\ \emph {et~al.}(2018)\citenamefont
  {Gresham}, \citenamefont {Lou},\ and\ \citenamefont
  {Zurek}}]{Gresham:2017cvl}%
  \BibitemOpen
  \bibfield  {author} {\bibinfo {author} {\bibfnamefont {M.~I.}\ \bibnamefont
  {Gresham}}, \bibinfo {author} {\bibfnamefont {H.~K.}\ \bibnamefont {Lou}}, \
  and\ \bibinfo {author} {\bibfnamefont {K.~M.}\ \bibnamefont {Zurek}},\ }\href
  {\doibase 10.1103/PhysRevD.97.036003} {\bibfield  {journal} {\bibinfo
  {journal} {Phys. Rev. D}\ }\textbf {\bibinfo {volume} {97}},\ \bibinfo
  {pages} {036003} (\bibinfo {year} {2018})},\ \Eprint
  {http://arxiv.org/abs/1707.02316} {arXiv:1707.02316 [hep-ph]} \BibitemShut
  {NoStop}%
\bibitem [{\citenamefont {Gresham}\ and\ \citenamefont
  {Zurek}(2019)}]{Gresham:2018rqo}%
  \BibitemOpen
  \bibfield  {author} {\bibinfo {author} {\bibfnamefont {M.~I.}\ \bibnamefont
  {Gresham}}\ and\ \bibinfo {author} {\bibfnamefont {K.~M.}\ \bibnamefont
  {Zurek}},\ }\href {\doibase 10.1103/PhysRevD.99.083008} {\bibfield  {journal}
  {\bibinfo  {journal} {Phys. Rev. D}\ }\textbf {\bibinfo {volume} {99}},\
  \bibinfo {pages} {083008} (\bibinfo {year} {2019})},\ \Eprint
  {http://arxiv.org/abs/1809.08254} {arXiv:1809.08254 [astro-ph.CO]}
  \BibitemShut {NoStop}%
\bibitem [{\citenamefont {Sakharov}(1967)}]{Sakharov:1967dj}%
  \BibitemOpen
  \bibfield  {author} {\bibinfo {author} {\bibfnamefont {A.~D.}\ \bibnamefont
  {Sakharov}},\ }\href {\doibase 10.1070/PU1991v034n05ABEH002497} {\bibfield
  {journal} {\bibinfo  {journal} {Pisma Zh. Eksp. Teor. Fiz.}\ }\textbf
  {\bibinfo {volume} {5}},\ \bibinfo {pages} {32} (\bibinfo {year}
  {1967})}\BibitemShut {NoStop}%
\bibitem [{\citenamefont {Kuzmin}\ \emph {et~al.}(1985)\citenamefont {Kuzmin},
  \citenamefont {Rubakov},\ and\ \citenamefont {Shaposhnikov}}]{Kuzmin:1985mm}%
  \BibitemOpen
  \bibfield  {author} {\bibinfo {author} {\bibfnamefont {V.~A.}\ \bibnamefont
  {Kuzmin}}, \bibinfo {author} {\bibfnamefont {V.~A.}\ \bibnamefont {Rubakov}},
  \ and\ \bibinfo {author} {\bibfnamefont {M.~E.}\ \bibnamefont
  {Shaposhnikov}},\ }\href {\doibase 10.1016/0370-2693(85)91028-7} {\bibfield
  {journal} {\bibinfo  {journal} {Phys. Lett. B}\ }\textbf {\bibinfo {volume}
  {155}},\ \bibinfo {pages} {36} (\bibinfo {year} {1985})}\BibitemShut
  {NoStop}%
\bibitem [{\citenamefont {Garcia-Bellido}\ \emph {et~al.}(1999)\citenamefont
  {Garcia-Bellido}, \citenamefont {Grigoriev}, \citenamefont {Kusenko},\ and\
  \citenamefont {Shaposhnikov}}]{Garcia-Bellido:1999xos}%
  \BibitemOpen
  \bibfield  {author} {\bibinfo {author} {\bibfnamefont {J.}~\bibnamefont
  {Garcia-Bellido}}, \bibinfo {author} {\bibfnamefont {D.~Y.}\ \bibnamefont
  {Grigoriev}}, \bibinfo {author} {\bibfnamefont {A.}~\bibnamefont {Kusenko}},
  \ and\ \bibinfo {author} {\bibfnamefont {M.~E.}\ \bibnamefont
  {Shaposhnikov}},\ }\href {\doibase 10.1103/PhysRevD.60.123504} {\bibfield
  {journal} {\bibinfo  {journal} {Phys. Rev. D}\ }\textbf {\bibinfo {volume}
  {60}},\ \bibinfo {pages} {123504} (\bibinfo {year} {1999})},\ \Eprint
  {http://arxiv.org/abs/hep-ph/9902449} {arXiv:hep-ph/9902449} \BibitemShut
  {NoStop}%
\bibitem [{\citenamefont {Cornwall}\ \emph {et~al.}(2001)\citenamefont
  {Cornwall}, \citenamefont {Grigoriev},\ and\ \citenamefont
  {Kusenko}}]{Cornwall:2001hq}%
  \BibitemOpen
  \bibfield  {author} {\bibinfo {author} {\bibfnamefont {J.~M.}\ \bibnamefont
  {Cornwall}}, \bibinfo {author} {\bibfnamefont {D.}~\bibnamefont {Grigoriev}},
  \ and\ \bibinfo {author} {\bibfnamefont {A.}~\bibnamefont {Kusenko}},\ }\href
  {\doibase 10.1103/PhysRevD.64.123518} {\bibfield  {journal} {\bibinfo
  {journal} {Phys. Rev. D}\ }\textbf {\bibinfo {volume} {64}},\ \bibinfo
  {pages} {123518} (\bibinfo {year} {2001})},\ \Eprint
  {http://arxiv.org/abs/hep-ph/0106127} {arXiv:hep-ph/0106127} \BibitemShut
  {NoStop}%
\bibitem [{\citenamefont {Asaka}\ \emph {et~al.}(2004)\citenamefont {Asaka},
  \citenamefont {Grigoriev}, \citenamefont {Kuzmin},\ and\ \citenamefont
  {Shaposhnikov}}]{Asaka:2003vt}%
  \BibitemOpen
  \bibfield  {author} {\bibinfo {author} {\bibfnamefont {T.}~\bibnamefont
  {Asaka}}, \bibinfo {author} {\bibfnamefont {D.}~\bibnamefont {Grigoriev}},
  \bibinfo {author} {\bibfnamefont {V.}~\bibnamefont {Kuzmin}}, \ and\ \bibinfo
  {author} {\bibfnamefont {M.}~\bibnamefont {Shaposhnikov}},\ }\href {\doibase
  10.1103/PhysRevLett.92.101303} {\bibfield  {journal} {\bibinfo  {journal}
  {Phys. Rev. Lett.}\ }\textbf {\bibinfo {volume} {92}},\ \bibinfo {pages}
  {101303} (\bibinfo {year} {2004})},\ \Eprint
  {http://arxiv.org/abs/hep-ph/0310100} {arXiv:hep-ph/0310100} \BibitemShut
  {NoStop}%
\bibitem [{\citenamefont {Braaten}\ and\ \citenamefont
  {Pisarski}(1990)}]{Braaten:1990it}%
  \BibitemOpen
  \bibfield  {author} {\bibinfo {author} {\bibfnamefont {E.}~\bibnamefont
  {Braaten}}\ and\ \bibinfo {author} {\bibfnamefont {R.~D.}\ \bibnamefont
  {Pisarski}},\ }\href {\doibase 10.1103/PhysRevD.42.2156} {\bibfield
  {journal} {\bibinfo  {journal} {Phys. Rev. D}\ }\textbf {\bibinfo {volume}
  {42}},\ \bibinfo {pages} {2156} (\bibinfo {year} {1990})}\BibitemShut
  {NoStop}%
\bibitem [{\citenamefont {Chowdhury}\ and\ \citenamefont
  {Eberhardt}(2018)}]{Chowdhury:2017aav}%
  \BibitemOpen
  \bibfield  {author} {\bibinfo {author} {\bibfnamefont {D.}~\bibnamefont
  {Chowdhury}}\ and\ \bibinfo {author} {\bibfnamefont {O.}~\bibnamefont
  {Eberhardt}},\ }\href {\doibase 10.1007/JHEP05(2018)161} {\bibfield
  {journal} {\bibinfo  {journal} {JHEP}\ }\textbf {\bibinfo {volume} {05}},\
  \bibinfo {pages} {161} (\bibinfo {year} {2018})},\ \Eprint
  {http://arxiv.org/abs/1711.02095} {arXiv:1711.02095 [hep-ph]} \BibitemShut
  {NoStop}%
\bibitem [{\citenamefont {Bhattacharyya}\ and\ \citenamefont
  {Das}(2016)}]{Bhattacharyya:2015nca}%
  \BibitemOpen
  \bibfield  {author} {\bibinfo {author} {\bibfnamefont {G.}~\bibnamefont
  {Bhattacharyya}}\ and\ \bibinfo {author} {\bibfnamefont {D.}~\bibnamefont
  {Das}},\ }\href {\doibase 10.1007/s12043-016-1252-4} {\bibfield  {journal}
  {\bibinfo  {journal} {Pramana}\ }\textbf {\bibinfo {volume} {87}},\ \bibinfo
  {pages} {40} (\bibinfo {year} {2016})},\ \Eprint
  {http://arxiv.org/abs/1507.06424} {arXiv:1507.06424 [hep-ph]} \BibitemShut
  {NoStop}%
\bibitem [{\citenamefont {Dine}\ \emph {et~al.}(1991)\citenamefont {Dine},
  \citenamefont {Huet}, \citenamefont {Singleton},\ and\ \citenamefont
  {Susskind}}]{Dine:1990fj}%
  \BibitemOpen
  \bibfield  {author} {\bibinfo {author} {\bibfnamefont {M.}~\bibnamefont
  {Dine}}, \bibinfo {author} {\bibfnamefont {P.}~\bibnamefont {Huet}}, \bibinfo
  {author} {\bibfnamefont {R.~L.}\ \bibnamefont {Singleton}, \bibfnamefont
  {Jr}}, \ and\ \bibinfo {author} {\bibfnamefont {L.}~\bibnamefont
  {Susskind}},\ }\href {\doibase 10.1016/0370-2693(91)91905-B} {\bibfield
  {journal} {\bibinfo  {journal} {Phys. Lett. B}\ }\textbf {\bibinfo {volume}
  {257}},\ \bibinfo {pages} {351} (\bibinfo {year} {1991})}\BibitemShut
  {NoStop}%
\bibitem [{\citenamefont {Cohen}\ \emph {et~al.}(1991)\citenamefont {Cohen},
  \citenamefont {Kaplan},\ and\ \citenamefont {Nelson}}]{Cohen:1990it}%
  \BibitemOpen
  \bibfield  {author} {\bibinfo {author} {\bibfnamefont {A.~G.}\ \bibnamefont
  {Cohen}}, \bibinfo {author} {\bibfnamefont {D.~B.}\ \bibnamefont {Kaplan}}, \
  and\ \bibinfo {author} {\bibfnamefont {A.~E.}\ \bibnamefont {Nelson}},\
  }\href {\doibase 10.1016/0550-3213(91)90395-E} {\bibfield  {journal}
  {\bibinfo  {journal} {Nucl. Phys. B}\ }\textbf {\bibinfo {volume} {349}},\
  \bibinfo {pages} {727} (\bibinfo {year} {1991})}\BibitemShut {NoStop}%
\bibitem [{\citenamefont {Dolgov}\ \emph {et~al.}(1997)\citenamefont {Dolgov},
  \citenamefont {Freese}, \citenamefont {Rangarajan},\ and\ \citenamefont
  {Srednicki}}]{Dolgov:1996qq}%
  \BibitemOpen
  \bibfield  {author} {\bibinfo {author} {\bibfnamefont {A.}~\bibnamefont
  {Dolgov}}, \bibinfo {author} {\bibfnamefont {K.}~\bibnamefont {Freese}},
  \bibinfo {author} {\bibfnamefont {R.}~\bibnamefont {Rangarajan}}, \ and\
  \bibinfo {author} {\bibfnamefont {M.}~\bibnamefont {Srednicki}},\ }\href
  {\doibase 10.1103/PhysRevD.56.6155} {\bibfield  {journal} {\bibinfo
  {journal} {Phys. Rev. D}\ }\textbf {\bibinfo {volume} {56}},\ \bibinfo
  {pages} {6155} (\bibinfo {year} {1997})},\ \Eprint
  {http://arxiv.org/abs/hep-ph/9610405} {arXiv:hep-ph/9610405} \BibitemShut
  {NoStop}%
\bibitem [{\citenamefont {Dolgov}(1997)}]{Dolgov:1997qr}%
  \BibitemOpen
  \bibfield  {author} {\bibinfo {author} {\bibfnamefont {A.~D.}\ \bibnamefont
  {Dolgov}},\ }in\ \href {\doibase 10.1080/01422419808240874} {\emph {\bibinfo
  {booktitle} {{25th ITEP Winter School of Physics}}}}\ (\bibinfo {year}
  {1997})\ \Eprint {http://arxiv.org/abs/hep-ph/9707419} {arXiv:hep-ph/9707419}
  \BibitemShut {NoStop}%
\bibitem [{\citenamefont {Kusenko}\ \emph
  {et~al.}(2015{\natexlab{a}})\citenamefont {Kusenko}, \citenamefont {Pearce},\
  and\ \citenamefont {Yang}}]{Kusenko:2014lra}%
  \BibitemOpen
  \bibfield  {author} {\bibinfo {author} {\bibfnamefont {A.}~\bibnamefont
  {Kusenko}}, \bibinfo {author} {\bibfnamefont {L.}~\bibnamefont {Pearce}}, \
  and\ \bibinfo {author} {\bibfnamefont {L.}~\bibnamefont {Yang}},\ }\href
  {\doibase 10.1103/PhysRevLett.114.061302} {\bibfield  {journal} {\bibinfo
  {journal} {Phys. Rev. Lett.}\ }\textbf {\bibinfo {volume} {114}},\ \bibinfo
  {pages} {061302} (\bibinfo {year} {2015}{\natexlab{a}})},\ \Eprint
  {http://arxiv.org/abs/1410.0722} {arXiv:1410.0722 [hep-ph]} \BibitemShut
  {NoStop}%
\bibitem [{\citenamefont {Kusenko}\ \emph
  {et~al.}(2015{\natexlab{b}})\citenamefont {Kusenko}, \citenamefont
  {Schmitz},\ and\ \citenamefont {Yanagida}}]{Kusenko:2014uta}%
  \BibitemOpen
  \bibfield  {author} {\bibinfo {author} {\bibfnamefont {A.}~\bibnamefont
  {Kusenko}}, \bibinfo {author} {\bibfnamefont {K.}~\bibnamefont {Schmitz}}, \
  and\ \bibinfo {author} {\bibfnamefont {T.~T.}\ \bibnamefont {Yanagida}},\
  }\href {\doibase 10.1103/PhysRevLett.115.011302} {\bibfield  {journal}
  {\bibinfo  {journal} {Phys. Rev. Lett.}\ }\textbf {\bibinfo {volume} {115}},\
  \bibinfo {pages} {011302} (\bibinfo {year} {2015}{\natexlab{b}})},\ \Eprint
  {http://arxiv.org/abs/1412.2043} {arXiv:1412.2043 [hep-ph]} \BibitemShut
  {NoStop}%
\bibitem [{\citenamefont {Yang}\ \emph {et~al.}(2015)\citenamefont {Yang},
  \citenamefont {Pearce},\ and\ \citenamefont {Kusenko}}]{Yang:2015ida}%
  \BibitemOpen
  \bibfield  {author} {\bibinfo {author} {\bibfnamefont {L.}~\bibnamefont
  {Yang}}, \bibinfo {author} {\bibfnamefont {L.}~\bibnamefont {Pearce}}, \ and\
  \bibinfo {author} {\bibfnamefont {A.}~\bibnamefont {Kusenko}},\ }\href
  {\doibase 10.1103/PhysRevD.92.043506} {\bibfield  {journal} {\bibinfo
  {journal} {Phys. Rev. D}\ }\textbf {\bibinfo {volume} {92}},\ \bibinfo
  {pages} {043506} (\bibinfo {year} {2015})},\ \Eprint
  {http://arxiv.org/abs/1505.07912} {arXiv:1505.07912 [hep-ph]} \BibitemShut
  {NoStop}%
\bibitem [{\citenamefont {Bodeker}\ \emph {et~al.}(2000)\citenamefont
  {Bodeker}, \citenamefont {Moore},\ and\ \citenamefont
  {Rummukainen}}]{Bodeker:1999gx}%
  \BibitemOpen
  \bibfield  {author} {\bibinfo {author} {\bibfnamefont {D.}~\bibnamefont
  {Bodeker}}, \bibinfo {author} {\bibfnamefont {G.~D.}\ \bibnamefont {Moore}},
  \ and\ \bibinfo {author} {\bibfnamefont {K.}~\bibnamefont {Rummukainen}},\
  }\href {\doibase 10.1103/PhysRevD.61.056003} {\bibfield  {journal} {\bibinfo
  {journal} {Phys. Rev. D}\ }\textbf {\bibinfo {volume} {61}},\ \bibinfo
  {pages} {056003} (\bibinfo {year} {2000})},\ \Eprint
  {http://arxiv.org/abs/hep-ph/9907545} {arXiv:hep-ph/9907545} \BibitemShut
  {NoStop}%
\bibitem [{\citenamefont {Moore}\ and\ \citenamefont
  {Rummukainen}(2000)}]{Moore:1999fs}%
  \BibitemOpen
  \bibfield  {author} {\bibinfo {author} {\bibfnamefont {G.~D.}\ \bibnamefont
  {Moore}}\ and\ \bibinfo {author} {\bibfnamefont {K.}~\bibnamefont
  {Rummukainen}},\ }\href {\doibase 10.1103/PhysRevD.61.105008} {\bibfield
  {journal} {\bibinfo  {journal} {Phys. Rev. D}\ }\textbf {\bibinfo {volume}
  {61}},\ \bibinfo {pages} {105008} (\bibinfo {year} {2000})},\ \Eprint
  {http://arxiv.org/abs/hep-ph/9906259} {arXiv:hep-ph/9906259} \BibitemShut
  {NoStop}%
\bibitem [{\citenamefont {Moore}(2000)}]{Moore:2000mx}%
  \BibitemOpen
  \bibfield  {author} {\bibinfo {author} {\bibfnamefont {G.~D.}\ \bibnamefont
  {Moore}},\ }\href {\doibase 10.1103/PhysRevD.62.085011} {\bibfield  {journal}
  {\bibinfo  {journal} {Phys. Rev. D}\ }\textbf {\bibinfo {volume} {62}},\
  \bibinfo {pages} {085011} (\bibinfo {year} {2000})},\ \Eprint
  {http://arxiv.org/abs/hep-ph/0001216} {arXiv:hep-ph/0001216} \BibitemShut
  {NoStop}%
\bibitem [{\citenamefont {Carson}\ \emph {et~al.}(1990)\citenamefont {Carson},
  \citenamefont {Li}, \citenamefont {McLerran},\ and\ \citenamefont
  {Wang}}]{Carson:1990jm}%
  \BibitemOpen
  \bibfield  {author} {\bibinfo {author} {\bibfnamefont {L.}~\bibnamefont
  {Carson}}, \bibinfo {author} {\bibfnamefont {X.}~\bibnamefont {Li}}, \bibinfo
  {author} {\bibfnamefont {L.~D.}\ \bibnamefont {McLerran}}, \ and\ \bibinfo
  {author} {\bibfnamefont {R.-T.}\ \bibnamefont {Wang}},\ }\href {\doibase
  10.1103/PhysRevD.42.2127} {\bibfield  {journal} {\bibinfo  {journal} {Phys.
  Rev. D}\ }\textbf {\bibinfo {volume} {42}},\ \bibinfo {pages} {2127}
  (\bibinfo {year} {1990})}\BibitemShut {NoStop}%
\bibitem [{\citenamefont {Baacke}\ and\ \citenamefont
  {Junker}(1994)}]{Baacke:1993aj}%
  \BibitemOpen
  \bibfield  {author} {\bibinfo {author} {\bibfnamefont {J.}~\bibnamefont
  {Baacke}}\ and\ \bibinfo {author} {\bibfnamefont {S.}~\bibnamefont
  {Junker}},\ }\href {\doibase 10.1103/PhysRevD.49.2055} {\bibfield  {journal}
  {\bibinfo  {journal} {Phys. Rev. D}\ }\textbf {\bibinfo {volume} {49}},\
  \bibinfo {pages} {2055} (\bibinfo {year} {1994})},\ \Eprint
  {http://arxiv.org/abs/hep-ph/9308310} {arXiv:hep-ph/9308310} \BibitemShut
  {NoStop}%
\bibitem [{\citenamefont {D'Onofrio}\ \emph {et~al.}(2014)\citenamefont
  {D'Onofrio}, \citenamefont {Rummukainen},\ and\ \citenamefont
  {Tranberg}}]{DOnofrio:2014rug}%
  \BibitemOpen
  \bibfield  {author} {\bibinfo {author} {\bibfnamefont {M.}~\bibnamefont
  {D'Onofrio}}, \bibinfo {author} {\bibfnamefont {K.}~\bibnamefont
  {Rummukainen}}, \ and\ \bibinfo {author} {\bibfnamefont {A.}~\bibnamefont
  {Tranberg}},\ }\href {\doibase 10.1103/PhysRevLett.113.141602} {\bibfield
  {journal} {\bibinfo  {journal} {Phys. Rev. Lett.}\ }\textbf {\bibinfo
  {volume} {113}},\ \bibinfo {pages} {141602} (\bibinfo {year} {2014})},\
  \Eprint {http://arxiv.org/abs/1404.3565} {arXiv:1404.3565 [hep-ph]}
  \BibitemShut {NoStop}%
\bibitem [{\citenamefont {Aghanim}\ \emph {et~al.}(2020)\citenamefont {Aghanim}
  \emph {et~al.}}]{Planck:2018vyg}%
  \BibitemOpen
  \bibfield  {author} {\bibinfo {author} {\bibfnamefont {N.}~\bibnamefont
  {Aghanim}} \emph {et~al.} (\bibinfo {collaboration} {Planck}),\ }\href
  {\doibase 10.1051/0004-6361/201833910} {\bibfield  {journal} {\bibinfo
  {journal} {Astron. Astrophys.}\ }\textbf {\bibinfo {volume} {641}},\ \bibinfo
  {pages} {A6} (\bibinfo {year} {2020})},\ \bibinfo {note} {[Erratum:
  Astron.Astrophys. 652, C4 (2021)]},\ \Eprint
  {http://arxiv.org/abs/1807.06209} {arXiv:1807.06209 [astro-ph.CO]}
  \BibitemShut {NoStop}%
\bibitem [{\citenamefont {Arnold}\ \emph {et~al.}(2000)\citenamefont {Arnold},
  \citenamefont {Moore},\ and\ \citenamefont {Yaffe}}]{Arnold:2000dr}%
  \BibitemOpen
  \bibfield  {author} {\bibinfo {author} {\bibfnamefont {P.~B.}\ \bibnamefont
  {Arnold}}, \bibinfo {author} {\bibfnamefont {G.~D.}\ \bibnamefont {Moore}}, \
  and\ \bibinfo {author} {\bibfnamefont {L.~G.}\ \bibnamefont {Yaffe}},\ }\href
  {\doibase 10.1088/1126-6708/2000/11/001} {\bibfield  {journal} {\bibinfo
  {journal} {JHEP}\ }\textbf {\bibinfo {volume} {11}},\ \bibinfo {pages} {001}
  (\bibinfo {year} {2000})},\ \Eprint {http://arxiv.org/abs/hep-ph/0010177}
  {arXiv:hep-ph/0010177} \BibitemShut {NoStop}%
\bibitem [{\citenamefont {Jedamzik}\ and\ \citenamefont
  {Rehm}(2001)}]{Jedamzik:2001qc}%
  \BibitemOpen
  \bibfield  {author} {\bibinfo {author} {\bibfnamefont {K.}~\bibnamefont
  {Jedamzik}}\ and\ \bibinfo {author} {\bibfnamefont {J.~B.}\ \bibnamefont
  {Rehm}},\ }\href {\doibase 10.1103/PhysRevD.64.023510} {\bibfield  {journal}
  {\bibinfo  {journal} {Phys. Rev. D}\ }\textbf {\bibinfo {volume} {64}},\
  \bibinfo {pages} {023510} (\bibinfo {year} {2001})},\ \Eprint
  {http://arxiv.org/abs/astro-ph/0101292} {arXiv:astro-ph/0101292} \BibitemShut
  {NoStop}%
\bibitem [{\citenamefont {Matsuura}\ \emph {et~al.}(2004)\citenamefont
  {Matsuura}, \citenamefont {Dolgov}, \citenamefont {Nagataki},\ and\
  \citenamefont {Sato}}]{Matsuura:2004ss}%
  \BibitemOpen
  \bibfield  {author} {\bibinfo {author} {\bibfnamefont {S.}~\bibnamefont
  {Matsuura}}, \bibinfo {author} {\bibfnamefont {A.~D.}\ \bibnamefont
  {Dolgov}}, \bibinfo {author} {\bibfnamefont {S.}~\bibnamefont {Nagataki}}, \
  and\ \bibinfo {author} {\bibfnamefont {K.}~\bibnamefont {Sato}},\ }\href
  {\doibase 10.1143/PTP.112.971} {\bibfield  {journal} {\bibinfo  {journal}
  {Prog. Theor. Phys.}\ }\textbf {\bibinfo {volume} {112}},\ \bibinfo {pages}
  {971} (\bibinfo {year} {2004})},\ \Eprint
  {http://arxiv.org/abs/astro-ph/0405459} {arXiv:astro-ph/0405459} \BibitemShut
  {NoStop}%
\bibitem [{\citenamefont {Dolgov}\ \emph {et~al.}(2009)\citenamefont {Dolgov},
  \citenamefont {Kawasaki},\ and\ \citenamefont {Kevlishvili}}]{Dolgov:2008wu}%
  \BibitemOpen
  \bibfield  {author} {\bibinfo {author} {\bibfnamefont {A.~D.}\ \bibnamefont
  {Dolgov}}, \bibinfo {author} {\bibfnamefont {M.}~\bibnamefont {Kawasaki}}, \
  and\ \bibinfo {author} {\bibfnamefont {N.}~\bibnamefont {Kevlishvili}},\
  }\href {\doibase 10.1016/j.nuclphysb.2008.08.029} {\bibfield  {journal}
  {\bibinfo  {journal} {Nucl. Phys. B}\ }\textbf {\bibinfo {volume} {807}},\
  \bibinfo {pages} {229} (\bibinfo {year} {2009})},\ \Eprint
  {http://arxiv.org/abs/0806.2986} {arXiv:0806.2986 [hep-ph]} \BibitemShut
  {NoStop}%
\bibitem [{\citenamefont {Nakamura}\ \emph {et~al.}(2017)\citenamefont
  {Nakamura}, \citenamefont {Hasahimoto}, \citenamefont {Ichimasa},\ and\
  \citenamefont {Arai}}]{Nakamura:2017qtu}%
  \BibitemOpen
  \bibfield  {author} {\bibinfo {author} {\bibfnamefont {R.}~\bibnamefont
  {Nakamura}}, \bibinfo {author} {\bibfnamefont {M.-a.}\ \bibnamefont
  {Hasahimoto}}, \bibinfo {author} {\bibfnamefont {R.}~\bibnamefont
  {Ichimasa}}, \ and\ \bibinfo {author} {\bibfnamefont {K.}~\bibnamefont
  {Arai}},\ }\href {\doibase 10.1142/S0218301317410038} {\bibfield  {journal}
  {\bibinfo  {journal} {Int. J. Mod. Phys. E}\ }\textbf {\bibinfo {volume}
  {26}},\ \bibinfo {pages} {1741003} (\bibinfo {year} {2017})},\ \Eprint
  {http://arxiv.org/abs/1710.08153} {arXiv:1710.08153 [astro-ph.CO]}
  \BibitemShut {NoStop}%
\bibitem [{\citenamefont {Blennow}\ \emph {et~al.}(2012)\citenamefont
  {Blennow}, \citenamefont {Fernandez-Martinez}, \citenamefont {Mena},
  \citenamefont {Redondo},\ and\ \citenamefont {Serra}}]{Blennow:2012de}%
  \BibitemOpen
  \bibfield  {author} {\bibinfo {author} {\bibfnamefont {M.}~\bibnamefont
  {Blennow}}, \bibinfo {author} {\bibfnamefont {E.}~\bibnamefont
  {Fernandez-Martinez}}, \bibinfo {author} {\bibfnamefont {O.}~\bibnamefont
  {Mena}}, \bibinfo {author} {\bibfnamefont {J.}~\bibnamefont {Redondo}}, \
  and\ \bibinfo {author} {\bibfnamefont {P.}~\bibnamefont {Serra}},\ }\href
  {\doibase 10.1088/1475-7516/2012/07/022} {\bibfield  {journal} {\bibinfo
  {journal} {JCAP}\ }\textbf {\bibinfo {volume} {1207}},\ \bibinfo {pages}
  {022} (\bibinfo {year} {2012})},\ \Eprint {http://arxiv.org/abs/1203.5803}
  {arXiv:1203.5803 [hep-ph]} \BibitemShut {NoStop}%
\bibitem [{\citenamefont {Fuller}\ \emph {et~al.}(2011)\citenamefont {Fuller},
  \citenamefont {Kishimoto},\ and\ \citenamefont {Kusenko}}]{Fuller:2011qy}%
  \BibitemOpen
  \bibfield  {author} {\bibinfo {author} {\bibfnamefont {G.~M.}\ \bibnamefont
  {Fuller}}, \bibinfo {author} {\bibfnamefont {C.~T.}\ \bibnamefont
  {Kishimoto}}, \ and\ \bibinfo {author} {\bibfnamefont {A.}~\bibnamefont
  {Kusenko}},\ }\href@noop {} {\  (\bibinfo {year} {2011})},\ \Eprint
  {http://arxiv.org/abs/1110.6479} {arXiv:1110.6479 [astro-ph.CO]} \BibitemShut
  {NoStop}%
\bibitem [{\citenamefont {Patwardhan}\ \emph {et~al.}(2015)\citenamefont
  {Patwardhan}, \citenamefont {Fuller}, \citenamefont {Kishimoto},\ and\
  \citenamefont {Kusenko}}]{Patwardhan:2015kga}%
  \BibitemOpen
  \bibfield  {author} {\bibinfo {author} {\bibfnamefont {A.~V.}\ \bibnamefont
  {Patwardhan}}, \bibinfo {author} {\bibfnamefont {G.~M.}\ \bibnamefont
  {Fuller}}, \bibinfo {author} {\bibfnamefont {C.~T.}\ \bibnamefont
  {Kishimoto}}, \ and\ \bibinfo {author} {\bibfnamefont {A.}~\bibnamefont
  {Kusenko}},\ }\href {\doibase 10.1103/PhysRevD.92.103509} {\bibfield
  {journal} {\bibinfo  {journal} {Phys. Rev. D}\ }\textbf {\bibinfo {volume}
  {92}},\ \bibinfo {pages} {103509} (\bibinfo {year} {2015})},\ \Eprint
  {http://arxiv.org/abs/1507.01977} {arXiv:1507.01977 [astro-ph.CO]}
  \BibitemShut {NoStop}%
\bibitem [{\citenamefont {Bernal}\ \emph {et~al.}(2016)\citenamefont {Bernal},
  \citenamefont {Verde},\ and\ \citenamefont {Riess}}]{Bernal:2016gxb}%
  \BibitemOpen
  \bibfield  {author} {\bibinfo {author} {\bibfnamefont {J.~L.}\ \bibnamefont
  {Bernal}}, \bibinfo {author} {\bibfnamefont {L.}~\bibnamefont {Verde}}, \
  and\ \bibinfo {author} {\bibfnamefont {A.~G.}\ \bibnamefont {Riess}},\ }\href
  {\doibase 10.1088/1475-7516/2016/10/019} {\bibfield  {journal} {\bibinfo
  {journal} {JCAP}\ }\textbf {\bibinfo {volume} {10}},\ \bibinfo {pages} {019}
  (\bibinfo {year} {2016})},\ \Eprint {http://arxiv.org/abs/1607.05617}
  {arXiv:1607.05617 [astro-ph.CO]} \BibitemShut {NoStop}%
\bibitem [{\citenamefont {Gelmini}\ \emph {et~al.}(2021)\citenamefont
  {Gelmini}, \citenamefont {Kusenko},\ and\ \citenamefont
  {Takhistov}}]{Gelmini:2019deq}%
  \BibitemOpen
  \bibfield  {author} {\bibinfo {author} {\bibfnamefont {G.~B.}\ \bibnamefont
  {Gelmini}}, \bibinfo {author} {\bibfnamefont {A.}~\bibnamefont {Kusenko}}, \
  and\ \bibinfo {author} {\bibfnamefont {V.}~\bibnamefont {Takhistov}},\ }\href
  {\doibase 10.1088/1475-7516/2021/06/002} {\bibfield  {journal} {\bibinfo
  {journal} {JCAP}\ }\textbf {\bibinfo {volume} {06}},\ \bibinfo {pages} {002}
  (\bibinfo {year} {2021})},\ \Eprint {http://arxiv.org/abs/1906.10136}
  {arXiv:1906.10136 [astro-ph.CO]} \BibitemShut {NoStop}%
\bibitem [{\citenamefont {Anchordoqui}\ and\ \citenamefont
  {Perez~Bergliaffa}(2019)}]{Anchordoqui:2019yzc}%
  \BibitemOpen
  \bibfield  {author} {\bibinfo {author} {\bibfnamefont {L.~A.}\ \bibnamefont
  {Anchordoqui}}\ and\ \bibinfo {author} {\bibfnamefont {S.~E.}\ \bibnamefont
  {Perez~Bergliaffa}},\ }\href {\doibase 10.1103/PhysRevD.100.123525}
  {\bibfield  {journal} {\bibinfo  {journal} {Phys. Rev. D}\ }\textbf {\bibinfo
  {volume} {100}},\ \bibinfo {pages} {123525} (\bibinfo {year} {2019})},\
  \Eprint {http://arxiv.org/abs/1910.05860} {arXiv:1910.05860 [astro-ph.CO]}
  \BibitemShut {NoStop}%
\bibitem [{\citenamefont {Vattis}\ \emph {et~al.}(2019)\citenamefont {Vattis},
  \citenamefont {Koushiappas},\ and\ \citenamefont {Loeb}}]{Vattis:2019efj}%
  \BibitemOpen
  \bibfield  {author} {\bibinfo {author} {\bibfnamefont {K.}~\bibnamefont
  {Vattis}}, \bibinfo {author} {\bibfnamefont {S.~M.}\ \bibnamefont
  {Koushiappas}}, \ and\ \bibinfo {author} {\bibfnamefont {A.}~\bibnamefont
  {Loeb}},\ }\href {\doibase 10.1103/PhysRevD.99.121302} {\bibfield  {journal}
  {\bibinfo  {journal} {Phys. Rev. D}\ }\textbf {\bibinfo {volume} {99}},\
  \bibinfo {pages} {121302} (\bibinfo {year} {2019})},\ \Eprint
  {http://arxiv.org/abs/1903.06220} {arXiv:1903.06220 [astro-ph.CO]}
  \BibitemShut {NoStop}%
\bibitem [{\citenamefont {Escudero}\ and\ \citenamefont
  {Witte}(2020)}]{Escudero:2019gvw}%
  \BibitemOpen
  \bibfield  {author} {\bibinfo {author} {\bibfnamefont {M.}~\bibnamefont
  {Escudero}}\ and\ \bibinfo {author} {\bibfnamefont {S.~J.}\ \bibnamefont
  {Witte}},\ }\href {\doibase 10.1140/epjc/s10052-020-7854-5} {\bibfield
  {journal} {\bibinfo  {journal} {Eur. Phys. J. C}\ }\textbf {\bibinfo {volume}
  {80}},\ \bibinfo {pages} {294} (\bibinfo {year} {2020})},\ \Eprint
  {http://arxiv.org/abs/1909.04044} {arXiv:1909.04044 [astro-ph.CO]}
  \BibitemShut {NoStop}%
\bibitem [{\citenamefont {Gelmini}\ \emph {et~al.}(2020)\citenamefont
  {Gelmini}, \citenamefont {Kawasaki}, \citenamefont {Kusenko}, \citenamefont
  {Murai},\ and\ \citenamefont {Takhistov}}]{Gelmini:2020ekg}%
  \BibitemOpen
  \bibfield  {author} {\bibinfo {author} {\bibfnamefont {G.~B.}\ \bibnamefont
  {Gelmini}}, \bibinfo {author} {\bibfnamefont {M.}~\bibnamefont {Kawasaki}},
  \bibinfo {author} {\bibfnamefont {A.}~\bibnamefont {Kusenko}}, \bibinfo
  {author} {\bibfnamefont {K.}~\bibnamefont {Murai}}, \ and\ \bibinfo {author}
  {\bibfnamefont {V.}~\bibnamefont {Takhistov}},\ }\href {\doibase
  10.1088/1475-7516/2020/09/051} {\bibfield  {journal} {\bibinfo  {journal}
  {JCAP}\ }\textbf {\bibinfo {volume} {09}},\ \bibinfo {pages} {051} (\bibinfo
  {year} {2020})},\ \Eprint {http://arxiv.org/abs/2005.06721} {arXiv:2005.06721
  [hep-ph]} \BibitemShut {NoStop}%
\bibitem [{\citenamefont {Vagnozzi}(2020)}]{Vagnozzi:2019ezj}%
  \BibitemOpen
  \bibfield  {author} {\bibinfo {author} {\bibfnamefont {S.}~\bibnamefont
  {Vagnozzi}},\ }\href {\doibase 10.1103/PhysRevD.102.023518} {\bibfield
  {journal} {\bibinfo  {journal} {Phys. Rev. D}\ }\textbf {\bibinfo {volume}
  {102}},\ \bibinfo {pages} {023518} (\bibinfo {year} {2020})},\ \Eprint
  {http://arxiv.org/abs/1907.07569} {arXiv:1907.07569 [astro-ph.CO]}
  \BibitemShut {NoStop}%
\bibitem [{\citenamefont {Wong}\ \emph {et~al.}(2020)\citenamefont {Wong} \emph
  {et~al.}}]{Wong:2019kwg}%
  \BibitemOpen
  \bibfield  {author} {\bibinfo {author} {\bibfnamefont {K.~C.}\ \bibnamefont
  {Wong}} \emph {et~al.},\ }\href {\doibase 10.1093/mnras/stz3094} {\bibfield
  {journal} {\bibinfo  {journal} {Mon. Not. Roy. Astron. Soc.}\ }\textbf
  {\bibinfo {volume} {498}},\ \bibinfo {pages} {1420} (\bibinfo {year}
  {2020})},\ \Eprint {http://arxiv.org/abs/1907.04869} {arXiv:1907.04869
  [astro-ph.CO]} \BibitemShut {NoStop}%
\bibitem [{\citenamefont {Perivolaropoulos}\ and\ \citenamefont
  {Skara}(2022)}]{Perivolaropoulos:2021jda}%
  \BibitemOpen
  \bibfield  {author} {\bibinfo {author} {\bibfnamefont {L.}~\bibnamefont
  {Perivolaropoulos}}\ and\ \bibinfo {author} {\bibfnamefont {F.}~\bibnamefont
  {Skara}},\ }\href {\doibase 10.1016/j.newar.2022.101659} {\bibfield
  {journal} {\bibinfo  {journal} {New Astron. Rev.}\ }\textbf {\bibinfo
  {volume} {95}} (\bibinfo {year} {2022}),\ 10.1016/j.newar.2022.101659},\
  \Eprint {http://arxiv.org/abs/2105.05208} {arXiv:2105.05208 [astro-ph.CO]}
  \BibitemShut {NoStop}%
\bibitem [{\citenamefont {Kawasaki}\ \emph {et~al.}(2000)\citenamefont
  {Kawasaki}, \citenamefont {Kohri},\ and\ \citenamefont
  {Sugiyama}}]{Kawasaki:2000en}%
  \BibitemOpen
  \bibfield  {author} {\bibinfo {author} {\bibfnamefont {M.}~\bibnamefont
  {Kawasaki}}, \bibinfo {author} {\bibfnamefont {K.}~\bibnamefont {Kohri}}, \
  and\ \bibinfo {author} {\bibfnamefont {N.}~\bibnamefont {Sugiyama}},\ }\href
  {\doibase 10.1103/PhysRevD.62.023506} {\bibfield  {journal} {\bibinfo
  {journal} {Phys. Rev. D}\ }\textbf {\bibinfo {volume} {62}},\ \bibinfo
  {pages} {023506} (\bibinfo {year} {2000})},\ \Eprint
  {http://arxiv.org/abs/astro-ph/0002127} {arXiv:astro-ph/0002127} \BibitemShut
  {NoStop}%
\bibitem [{\citenamefont {Hannestad}(2004)}]{Hannestad:2004px}%
  \BibitemOpen
  \bibfield  {author} {\bibinfo {author} {\bibfnamefont {S.}~\bibnamefont
  {Hannestad}},\ }\href {\doibase 10.1103/PhysRevD.70.043506} {\bibfield
  {journal} {\bibinfo  {journal} {Phys. Rev. D}\ }\textbf {\bibinfo {volume}
  {70}},\ \bibinfo {pages} {043506} (\bibinfo {year} {2004})},\ \Eprint
  {http://arxiv.org/abs/astro-ph/0403291} {arXiv:astro-ph/0403291} \BibitemShut
  {NoStop}%
\bibitem [{\citenamefont {Gelmini}\ \emph {et~al.}(2004)\citenamefont
  {Gelmini}, \citenamefont {Palomares-Ruiz},\ and\ \citenamefont
  {Pascoli}}]{Gelmini:2004ah}%
  \BibitemOpen
  \bibfield  {author} {\bibinfo {author} {\bibfnamefont {G.}~\bibnamefont
  {Gelmini}}, \bibinfo {author} {\bibfnamefont {S.}~\bibnamefont
  {Palomares-Ruiz}}, \ and\ \bibinfo {author} {\bibfnamefont {S.}~\bibnamefont
  {Pascoli}},\ }\href {\doibase 10.1103/PhysRevLett.93.081302} {\bibfield
  {journal} {\bibinfo  {journal} {Phys. Rev. Lett.}\ }\textbf {\bibinfo
  {volume} {93}},\ \bibinfo {pages} {081302} (\bibinfo {year} {2004})},\
  \Eprint {http://arxiv.org/abs/astro-ph/0403323} {arXiv:astro-ph/0403323}
  \BibitemShut {NoStop}%
\bibitem [{\citenamefont {Gelmini}\ \emph {et~al.}(2007)\citenamefont
  {Gelmini}, \citenamefont {Gondolo}, \citenamefont {Soldatenko},\ and\
  \citenamefont {Yaguna}}]{Gelmini:2006mr}%
  \BibitemOpen
  \bibfield  {author} {\bibinfo {author} {\bibfnamefont {G.~B.}\ \bibnamefont
  {Gelmini}}, \bibinfo {author} {\bibfnamefont {P.}~\bibnamefont {Gondolo}},
  \bibinfo {author} {\bibfnamefont {A.}~\bibnamefont {Soldatenko}}, \ and\
  \bibinfo {author} {\bibfnamefont {C.~E.}\ \bibnamefont {Yaguna}},\ }\href
  {\doibase 10.1103/PhysRevD.76.015010} {\bibfield  {journal} {\bibinfo
  {journal} {Phys. Rev. D}\ }\textbf {\bibinfo {volume} {76}},\ \bibinfo
  {pages} {015010} (\bibinfo {year} {2007})},\ \Eprint
  {http://arxiv.org/abs/hep-ph/0610379} {arXiv:hep-ph/0610379} \BibitemShut
  {NoStop}%
\bibitem [{\citenamefont {Grin}\ \emph {et~al.}(2008)\citenamefont {Grin},
  \citenamefont {Smith},\ and\ \citenamefont {Kamionkowski}}]{Grin:2007yg}%
  \BibitemOpen
  \bibfield  {author} {\bibinfo {author} {\bibfnamefont {D.}~\bibnamefont
  {Grin}}, \bibinfo {author} {\bibfnamefont {T.~L.}\ \bibnamefont {Smith}}, \
  and\ \bibinfo {author} {\bibfnamefont {M.}~\bibnamefont {Kamionkowski}},\
  }\href {\doibase 10.1103/PhysRevD.77.085020} {\bibfield  {journal} {\bibinfo
  {journal} {Phys. Rev. D}\ }\textbf {\bibinfo {volume} {77}},\ \bibinfo
  {pages} {085020} (\bibinfo {year} {2008})},\ \Eprint
  {http://arxiv.org/abs/0711.1352} {arXiv:0711.1352 [astro-ph]} \BibitemShut
  {NoStop}%
\bibitem [{\citenamefont {Gelmini}\ \emph {et~al.}(2008)\citenamefont
  {Gelmini}, \citenamefont {Osoba}, \citenamefont {Palomares-Ruiz},\ and\
  \citenamefont {Pascoli}}]{Gelmini:2008fq}%
  \BibitemOpen
  \bibfield  {author} {\bibinfo {author} {\bibfnamefont {G.}~\bibnamefont
  {Gelmini}}, \bibinfo {author} {\bibfnamefont {E.}~\bibnamefont {Osoba}},
  \bibinfo {author} {\bibfnamefont {S.}~\bibnamefont {Palomares-Ruiz}}, \ and\
  \bibinfo {author} {\bibfnamefont {S.}~\bibnamefont {Pascoli}},\ }\href
  {\doibase 10.1088/1475-7516/2008/10/029} {\bibfield  {journal} {\bibinfo
  {journal} {JCAP}\ }\textbf {\bibinfo {volume} {0810}},\ \bibinfo {pages}
  {029} (\bibinfo {year} {2008})},\ \Eprint {http://arxiv.org/abs/0803.2735}
  {arXiv:0803.2735 [astro-ph]} \BibitemShut {NoStop}%
\bibitem [{\citenamefont {Gelmini}(2008)}]{Gelmini:2008ti}%
  \BibitemOpen
  \bibfield  {author} {\bibinfo {author} {\bibfnamefont {G.~B.}\ \bibnamefont
  {Gelmini}},\ }in\ \href@noop {} {\emph {\bibinfo {booktitle} {{43rd
  Rencontres de Moriond on QCD and High Energy Interactions}}}}\ (\bibinfo
  {year} {2008})\ pp.\ \bibinfo {pages} {345--348},\ \Eprint
  {http://arxiv.org/abs/0805.1824} {arXiv:0805.1824 [hep-ph]} \BibitemShut
  {NoStop}%
\bibitem [{\citenamefont {Rehagen}\ and\ \citenamefont
  {Gelmini}(2015)}]{Rehagen:2015zma}%
  \BibitemOpen
  \bibfield  {author} {\bibinfo {author} {\bibfnamefont {T.}~\bibnamefont
  {Rehagen}}\ and\ \bibinfo {author} {\bibfnamefont {G.~B.}\ \bibnamefont
  {Gelmini}},\ }\href {\doibase 10.1088/1475-7516/2015/06/039} {\bibfield
  {journal} {\bibinfo  {journal} {JCAP}\ }\textbf {\bibinfo {volume} {06}},\
  \bibinfo {pages} {039} (\bibinfo {year} {2015})},\ \Eprint
  {http://arxiv.org/abs/1504.03768} {arXiv:1504.03768 [hep-ph]} \BibitemShut
  {NoStop}%
\bibitem [{\citenamefont {Affleck}\ and\ \citenamefont
  {Dine}(1985)}]{Affleck:1984fy}%
  \BibitemOpen
  \bibfield  {author} {\bibinfo {author} {\bibfnamefont {I.}~\bibnamefont
  {Affleck}}\ and\ \bibinfo {author} {\bibfnamefont {M.}~\bibnamefont {Dine}},\
  }\href {\doibase 10.1016/0550-3213(85)90021-5} {\bibfield  {journal}
  {\bibinfo  {journal} {Nucl. Phys.}\ }\textbf {\bibinfo {volume} {B249}},\
  \bibinfo {pages} {361} (\bibinfo {year} {1985})}\BibitemShut {NoStop}%
\bibitem [{\citenamefont {Dine}\ and\ \citenamefont
  {Kusenko}(2003)}]{Dine:2003ax}%
  \BibitemOpen
  \bibfield  {author} {\bibinfo {author} {\bibfnamefont {M.}~\bibnamefont
  {Dine}}\ and\ \bibinfo {author} {\bibfnamefont {A.}~\bibnamefont {Kusenko}},\
  }\href {\doibase 10.1103/RevModPhys.76.1} {\bibfield  {journal} {\bibinfo
  {journal} {Rev. Mod. Phys.}\ }\textbf {\bibinfo {volume} {76}},\ \bibinfo
  {pages} {1} (\bibinfo {year} {2003})},\ \Eprint
  {http://arxiv.org/abs/hep-ph/0303065} {arXiv:hep-ph/0303065 [hep-ph]}
  \BibitemShut {NoStop}%
\bibitem [{\citenamefont {Elor}\ and\ \citenamefont
  {McGehee}(2021)}]{Elor:2020tkc}%
  \BibitemOpen
  \bibfield  {author} {\bibinfo {author} {\bibfnamefont {G.}~\bibnamefont
  {Elor}}\ and\ \bibinfo {author} {\bibfnamefont {R.}~\bibnamefont {McGehee}},\
  }\href {\doibase 10.1103/PhysRevD.103.035005} {\bibfield  {journal} {\bibinfo
   {journal} {Phys. Rev. D}\ }\textbf {\bibinfo {volume} {103}},\ \bibinfo
  {pages} {035005} (\bibinfo {year} {2021})},\ \Eprint
  {http://arxiv.org/abs/2011.06115} {arXiv:2011.06115 [hep-ph]} \BibitemShut
  {NoStop}%
\bibitem [{\citenamefont {Elahi}\ \emph {et~al.}(2022)\citenamefont {Elahi},
  \citenamefont {Elor},\ and\ \citenamefont {McGehee}}]{Elahi:2021jia}%
  \BibitemOpen
  \bibfield  {author} {\bibinfo {author} {\bibfnamefont {F.}~\bibnamefont
  {Elahi}}, \bibinfo {author} {\bibfnamefont {G.}~\bibnamefont {Elor}}, \ and\
  \bibinfo {author} {\bibfnamefont {R.}~\bibnamefont {McGehee}},\ }\href
  {\doibase 10.1103/PhysRevD.105.055024} {\bibfield  {journal} {\bibinfo
  {journal} {Phys. Rev. D}\ }\textbf {\bibinfo {volume} {105}},\ \bibinfo
  {pages} {055024} (\bibinfo {year} {2022})},\ \Eprint
  {http://arxiv.org/abs/2109.09751} {arXiv:2109.09751 [hep-ph]} \BibitemShut
  {NoStop}%
\bibitem [{\citenamefont {Jaeckel}\ and\ \citenamefont
  {Yin}(2023)}]{Jaeckel:2022osh}%
  \BibitemOpen
  \bibfield  {author} {\bibinfo {author} {\bibfnamefont {J.}~\bibnamefont
  {Jaeckel}}\ and\ \bibinfo {author} {\bibfnamefont {W.}~\bibnamefont {Yin}},\
  }\href {\doibase 10.1103/PhysRevD.107.015001} {\bibfield  {journal} {\bibinfo
   {journal} {Phys. Rev. D}\ }\textbf {\bibinfo {volume} {107}},\ \bibinfo
  {pages} {015001} (\bibinfo {year} {2023})},\ \Eprint
  {http://arxiv.org/abs/2206.06376} {arXiv:2206.06376 [hep-ph]} \BibitemShut
  {NoStop}%
\bibitem [{\citenamefont {Batell}\ \emph {et~al.}(2009)\citenamefont {Batell},
  \citenamefont {Pospelov},\ and\ \citenamefont {Ritz}}]{Batell:2009di}%
  \BibitemOpen
  \bibfield  {author} {\bibinfo {author} {\bibfnamefont {B.}~\bibnamefont
  {Batell}}, \bibinfo {author} {\bibfnamefont {M.}~\bibnamefont {Pospelov}}, \
  and\ \bibinfo {author} {\bibfnamefont {A.}~\bibnamefont {Ritz}},\ }\href
  {\doibase 10.1103/PhysRevD.80.095024} {\bibfield  {journal} {\bibinfo
  {journal} {Phys. Rev. D}\ }\textbf {\bibinfo {volume} {80}},\ \bibinfo
  {pages} {095024} (\bibinfo {year} {2009})},\ \Eprint
  {http://arxiv.org/abs/0906.5614} {arXiv:0906.5614 [hep-ph]} \BibitemShut
  {NoStop}%
\bibitem [{\citenamefont {Kanemura}\ \emph {et~al.}(2010)\citenamefont
  {Kanemura}, \citenamefont {Matsumoto}, \citenamefont {Nabeshima},\ and\
  \citenamefont {Okada}}]{Kanemura:2010sh}%
  \BibitemOpen
  \bibfield  {author} {\bibinfo {author} {\bibfnamefont {S.}~\bibnamefont
  {Kanemura}}, \bibinfo {author} {\bibfnamefont {S.}~\bibnamefont {Matsumoto}},
  \bibinfo {author} {\bibfnamefont {T.}~\bibnamefont {Nabeshima}}, \ and\
  \bibinfo {author} {\bibfnamefont {N.}~\bibnamefont {Okada}},\ }\href
  {\doibase 10.1103/PhysRevD.82.055026} {\bibfield  {journal} {\bibinfo
  {journal} {Phys. Rev. D}\ }\textbf {\bibinfo {volume} {82}},\ \bibinfo
  {pages} {055026} (\bibinfo {year} {2010})},\ \Eprint
  {http://arxiv.org/abs/1005.5651} {arXiv:1005.5651 [hep-ph]} \BibitemShut
  {NoStop}%
\bibitem [{\citenamefont {Fichet}(2018)}]{Fichet:2017bng}%
  \BibitemOpen
  \bibfield  {author} {\bibinfo {author} {\bibfnamefont {S.}~\bibnamefont
  {Fichet}},\ }\href {\doibase 10.1103/PhysRevLett.120.131801} {\bibfield
  {journal} {\bibinfo  {journal} {Phys. Rev. Lett.}\ }\textbf {\bibinfo
  {volume} {120}},\ \bibinfo {pages} {131801} (\bibinfo {year} {2018})},\
  \Eprint {http://arxiv.org/abs/1705.10331} {arXiv:1705.10331 [hep-ph]}
  \BibitemShut {NoStop}%
\bibitem [{\citenamefont {Contino}\ \emph {et~al.}(2021)\citenamefont
  {Contino}, \citenamefont {Max},\ and\ \citenamefont
  {Mishra}}]{Contino:2020tix}%
  \BibitemOpen
  \bibfield  {author} {\bibinfo {author} {\bibfnamefont {R.}~\bibnamefont
  {Contino}}, \bibinfo {author} {\bibfnamefont {K.}~\bibnamefont {Max}}, \ and\
  \bibinfo {author} {\bibfnamefont {R.~K.}\ \bibnamefont {Mishra}},\ }\href
  {\doibase 10.1007/JHEP06(2021)127} {\bibfield  {journal} {\bibinfo  {journal}
  {JHEP}\ }\textbf {\bibinfo {volume} {06}},\ \bibinfo {pages} {127} (\bibinfo
  {year} {2021})},\ \Eprint {http://arxiv.org/abs/2012.08537} {arXiv:2012.08537
  [hep-ph]} \BibitemShut {NoStop}%
\bibitem [{\citenamefont {Laine}(1995)}]{Laine:1994zq}%
  \BibitemOpen
  \bibfield  {author} {\bibinfo {author} {\bibfnamefont {M.}~\bibnamefont
  {Laine}},\ }\href {\doibase 10.1103/PhysRevD.51.4525} {\bibfield  {journal}
  {\bibinfo  {journal} {Phys. Rev. D}\ }\textbf {\bibinfo {volume} {51}},\
  \bibinfo {pages} {4525} (\bibinfo {year} {1995})},\ \Eprint
  {http://arxiv.org/abs/hep-ph/9411252} {arXiv:hep-ph/9411252} \BibitemShut
  {NoStop}%
\bibitem [{\citenamefont {Farakos}\ \emph {et~al.}(1995)\citenamefont
  {Farakos}, \citenamefont {Kajantie}, \citenamefont {Rummukainen},\ and\
  \citenamefont {Shaposhnikov}}]{Farakos:1994xh}%
  \BibitemOpen
  \bibfield  {author} {\bibinfo {author} {\bibfnamefont {K.}~\bibnamefont
  {Farakos}}, \bibinfo {author} {\bibfnamefont {K.}~\bibnamefont {Kajantie}},
  \bibinfo {author} {\bibfnamefont {K.}~\bibnamefont {Rummukainen}}, \ and\
  \bibinfo {author} {\bibfnamefont {M.~E.}\ \bibnamefont {Shaposhnikov}},\
  }\href {\doibase 10.1016/0550-3213(95)80129-4} {\bibfield  {journal}
  {\bibinfo  {journal} {Nucl. Phys. B}\ }\textbf {\bibinfo {volume} {442}},\
  \bibinfo {pages} {317} (\bibinfo {year} {1995})},\ \Eprint
  {http://arxiv.org/abs/hep-lat/9412091} {arXiv:hep-lat/9412091} \BibitemShut
  {NoStop}%
\bibitem [{\citenamefont {Niemi}\ \emph {et~al.}(2021)\citenamefont {Niemi},
  \citenamefont {Ramsey-Musolf}, \citenamefont {Tenkanen},\ and\ \citenamefont
  {Weir}}]{Niemi:2020hto}%
  \BibitemOpen
  \bibfield  {author} {\bibinfo {author} {\bibfnamefont {L.}~\bibnamefont
  {Niemi}}, \bibinfo {author} {\bibfnamefont {M.~J.}\ \bibnamefont
  {Ramsey-Musolf}}, \bibinfo {author} {\bibfnamefont {T.~V.~I.}\ \bibnamefont
  {Tenkanen}}, \ and\ \bibinfo {author} {\bibfnamefont {D.~J.}\ \bibnamefont
  {Weir}},\ }\href {\doibase 10.1103/PhysRevLett.126.171802} {\bibfield
  {journal} {\bibinfo  {journal} {Phys. Rev. Lett.}\ }\textbf {\bibinfo
  {volume} {126}},\ \bibinfo {pages} {171802} (\bibinfo {year} {2021})},\
  \Eprint {http://arxiv.org/abs/2005.11332} {arXiv:2005.11332 [hep-ph]}
  \BibitemShut {NoStop}%
\end{thebibliography}%

\appendix 
\subsection{One-Loop and Finite Temperature Corrections}

Our tree-level potential is given in Eq.~\eqref{eq:tree_potential}.  To study the potential at finite temperature we include one-loop Coleman-Weinberg corrections,
\begin{align}
V_{1-\mathrm{loop}} &=
\dfrac{M_1^5}{64\pi^2}
 \left(\log \left(\frac{M_1^2}{\phi_1^2 + \phi_2^2}\right)-\frac{3}{2}\right)
 \nonumber \\
& +\frac{M_2^4}{64 \pi ^2} \left(\log \left(\frac{M_2^2}{\phi_1^2 + \phi_2^2}\right)-\frac{3}{2}\right) \nonumber \\
& +\frac{6 M_W^4}{64 \pi ^2}
\left(\log \left(\frac{M_W^2}{\phi_1^2 + \phi_2^2}\right)-\frac{5}{6}\right) \nonumber \\
& +\frac{3 M_Z^4}{64 \pi ^2}  \left(\log \left(\frac{M_Z^2}{\phi_1^2 + \phi_2^2}\right)-\frac{5}{6}\right) \nonumber \\
& -\frac{12}{64 \pi ^2} \left(\frac{y_t^2 \phi_2^2 }{2}\right)^2 \left(\log \left(\frac{y_t^2  \phi_2^2}{2 \left(\phi_1^2 + \phi_2^2\right)}\right)-\frac{3}{2}\right)
\end{align}
where we have coupled the top quark only to the $H_2$ Higgs doublet, and $M_1$ and $M_2$ are the Higgs mass eigenvalues (calculated by diagonalizing the tree-level mass matrix).  We evaluate the gauge boson masses with the VEVs $\phi_1$ and $\phi_2$. 

We also include the finite temperature corrections
\begin{align}
V_T &= 
\frac{T^4}{2 \pi ^2} \left(J_B\left(\frac{M_1^2}{T^2}\right)
+J_B\left(\frac{M_2^2}{T^2}\right)+6 J_B\left(\frac{M_W^2}{T^2}\right) \right. \nonumber \\
& \qquad \left. +3 J_B\left(\frac{M_Z^2}{T^2}\right)-12 J_F\left(\frac{y_t^2 \phi_2^2}{2 T^2}\right)\right),
\end{align}
where the functions $J_B$ and $J_F$ are 
\begin{align}
J_{B,F}(y) &=
\int_0^\infty x^2 \ln \left( 1 \mp e^{-x^2 +y} \right).
\end{align}
Note that we do {\it not} include the leading order terms from resummation as, for a second order transition such as what we are considering, these terms actually make the infrared problem worse as $m_H^2 + \Pi$ vanishes at the origin near the critical temperature. This leads to the well-known problem of perturbation theory badly describing smooth transitions and misidentifying them as first order transitions \cite{Laine:1994zq,Farakos:1994xh,Niemi:2020hto}. Perturbation theory struggles to capture weak or second order transitions and as such our numerical results should be taken with a grain of salt indicating the qualitative dependence of the model on the parameters. \par
We considered a benchmark scenario with $\lambda_1 = 0.01844 $, $\lambda_2 =0.1676$, $ \lambda_3 =0.9513$, $\lambda_5 =0.8934$,
$ \lambda_6 =0.1149$,
$\xi =0 $, $v_1=77.5484 \, \mathrm{GeV}$, and $ v_2 = 234.9 \, \mathrm{GeV}$.  The mass eigenvalues of $V = V_{\rm tree} + V_{1-\rm{loop}}$, calculated at zero temperature, are $125$ and $378 \, \mathrm{GeV}$, and the minimum at zero temperature has $\sqrt{ \phi_1^2 + \phi_2^2} = 246.24 \, \mathrm{GeV}$.  Observational constraints are presented as functions of $\alpha = m_{12} \slash (m_{11} + m_{22})$, where $m_{ij}$ are elements of the mass matrix, and $\beta = \phi_2 \slash \phi_1$.  The values of these parameters for the zero-temperature minimum are given in the text.

\end{document}